\documentclass[aps,prl,floatfix,onecolumn,reprint,amsmath,amssymb,superscriptaddress]{revtex4-1}
\usepackage{amsfonts}
\usepackage{mathrsfs}
\usepackage{amsmath}
\usepackage{color}
\usepackage{natbib}
\usepackage{graphicx}
\usepackage{bm}
\usepackage{amssymb}
\usepackage{mathptmx}
\usepackage{stmaryrd}
\usepackage{xspace}
\usepackage{epstopdf}
\usepackage{dcolumn}
\usepackage{longtable}
\usepackage{multirow}
\usepackage{bbding}
\usepackage[colorlinks=true, letterpaper=true, pdfstartview=FitV, linkcolor=blue, citecolor=blue, urlcolor=blue]{hyperref}

\makeatletter

\newcommand{\Rmnum}[1]{\expandafter\@slowromancap\romannumeral #1@}

\makeatother
\begin{document}

\title{Phononic Obstructed Atomic Insulators with Robust Corner Modes 
}

\author{Da-Shuai Ma}
\affiliation{Institute for Structure and Function \& Department of Physics \& Chongqing Key Laboratory for Strongly Coupled Physics, Chongqing University, Chongqing 400044, P. R. China}
\affiliation{Center of Quantum materials and devices, Chongqing University, Chongqing 400044, P. R. China}

\author{Kejun Yu}
\affiliation{Centre for Quantum Physics, Key Laboratory of Advanced Optoelectronic Quantum Architecture and Measurement (MOE), School of Physics, Beijing Institute of Technology, Beijing 100081, China}
\affiliation{Beijing Key Lab of Nanophotonics and Ultrafine Optoelectronic Systems, School of Physics, Beijing Institute of Technology, Beijing 100081, China}

\author{Xiao-Ping Li}
\affiliation{School of Physical Science and Technology, Inner Mongolia University, Hohhot 010021, China}

\author{Xiaoyuan Zhou}
\email{xiaoyuan2013@cqu.edu.cn}
\affiliation{Center of Quantum materials and devices, Chongqing University, Chongqing 400044, P. R. China}

\author{Rui Wang}
\email{rcwang@cqu.edu.cn}
\affiliation{Institute for Structure and Function \& Department of Physics \& Chongqing Key Laboratory for Strongly Coupled Physics, Chongqing University, Chongqing 400044, P. R. China}
\affiliation{Center of Quantum materials and devices, Chongqing University, Chongqing 400044, P. R. China}

\begin{abstract}
Higher-order topological insulators (HOTIs) are described by symmetric exponentially decayed Wannier functions at some $necessary$ unoccupied  Wyckoff positions and classified as obstructed atomic insulators (OAIs) in the topological quantum chemistry (TQC) theory.
The boundary states in HOTIs reported so far are often fragile, manifested as strongly depending on crystalline symmetries and cleavage terminations in the disk or cylinder geometry.
Here, using the TQC theory, we present an intuitive argument about the connection between the obstructed Wannier charge centers of OAIs and the emergence of robust corner states in two-dimensional systems.
Based on first-principles calculations and Real Space Invariant theory, we extend the concept of OAIs to phonon systems and thereby predict that the robust corner states can be realized in the phonon spectra of $MX_3$ ($M$=Bi, Sb, As, Sc, Y; $X$=I, Br, Cl) monolayers.
The phonon corner modes in different shapes of nano-disks are investigated, and their robustness facilitates the detection in experiments and further applications.
This work suggests a promising avenue to explore more attractive features of higher-order band topology.
\end{abstract}

\maketitle

\textit{Introduction. ---}
The extension of quantized electric polarization from dipole moments to multipole moments has lead to intense recent studies of higher-order topological insulators (HOTIs) \cite{benalcazar2017quantized}.
The topological feature of HOTIs is characterized by nontrivial boundary states whose dimensionality is more than one below that of the bulk \cite{peterson2018quantized,schindler2018higher,PhysRevLett.119.246402,PhysRevLett.119.246401,schindler2018higher,PhysRevB.97.205136,PhysRevLett.123.186401,PhysRevB.99.245151,
PhysRevLett.122.256402,PhysRevLett.123.256402,PhysRevLett.124.136407,PhysRevLett.125.056402}, \textit{i.e.}, a HOTI in $d$ spatial dimensions possesses gapless boundary states at its  $(d-n)$-dimensional boundaries ($n>1$).
The brand-new bulk-boundary correspondence of HOTIs has enriched the community of band topology and thus attracted intensive attentions.
Among various phases of HOTIs, of particular importance is the two-dimensional (2D) second-order topological insulator (SOTI) where spatial distribution of ingap modes is localized at the corners of its 0D nano-disk~\cite{PhysRevLett.122.204301,PhysRevLett.122.233902,PhysRevLett.125.056402,lee2020two,PhysRevB.104.245427,mu2022kekule,PhysRevLett.128.026801}.
In sharp contrast to the helical surface states in traditional topological insulators (TIs) \cite{RevModPhys.82.3045}, these characteristic corner modes are separated from the conduction bands and valence bands with the corresponding energy level is adjustable by surface potential,
leaving more freedoms to detect them by spectroscopy experiments or be potentially applied in low dimensional devices.

While relevant advancements have been very encouraging, the presence of corner states in 2D SOTI reported so far strongly depends on the clipping geometries (\textit{i.e.}, crystalline symmetries and terminated atoms of nano-disks) \cite{PhysRevLett.120.026801,PhysRevLett.121.116801,PhysRevB.103.205123,PhysRevLett.123.256402,PhysRevLett.125.056402,wang2022two}, thus making that it is difficult for their experimental measurements or further practical applications.
As corner states are the most interesting features in 2D SOTIs, the exploration of these materials with robust corner states that are independent of the clipping forms is desirable.
Different from finding traditional first-order topological materials, the systematic discovery of HOTIs is in its infancy.
Fortunately, the rapid development of topological quantum chemistry (TQC) theory \cite{bradlyn2017topological,PhysRevB.97.035139} offers a reliable avenue to capture the higher-order band topology.
In the TQC theory, a system with non-zero Real Space Invariant (RSI) \cite{song2020twisted,xu2021three} at an empty Wyckoff positions (WPs) while the band representation (BR) is a sum of elementary band representations (EBRs) is a HOTI \cite{gao2022unconventional,xu2021three,guo2022quadrupole}.
Equivalently, the HOTI can be diagnosed by checking whether the system is an obstructed atomic insulator (OAI) where the obstructed Wannier charge centers (OWCCs) mismatch with the occupied WPs.
Furthermore, when a 2D OAI is clipped into 0D nano-disk where the OWCCs are exposed, the ingap corner states would be present accordingly.

On the other hand, it is worth noting that the spin-orbit coupling is $not$ a necessary condition for the presence of OAIs, indicating that the OAI can be realized in bosonic systems, such as photonic systems~\cite{serra2018observation,mittal2019photonic,el2019corner,PhysRevLett.122.233903}, acoustic systems~\cite{ma2019topological,xue2019acoustic,ni2019observation,PhysRevLett.124.206601}, and electric circuit~\cite{imhof2018topolectrical,PhysRevB.98.201402,PhysRevB.99.020304}.
Therefore, compared that the first-order phononic topological insulator is nearly impossible to occur in realistic materials, it is expected to find phononic OAIs with HOTI features.
In fact, the phonon in crystalline solids also works as another fancy platform for exploring nontrivial band topology and the related applications~\cite{PhysRevLett.105.225901,PhysRevLett.119.255901,PhysRevLett.120.016401,PhysRevLett.121.035302,PhysRevLett.123.245302,
PhysRevLett.123.065501,PhysRevLett.124.185501,liu2020topological,PhysRevLett.124.105303,PhysRevB.101.081403,li2021computation,PhysRevB.105.085123}.
However, the OAI and its fascinating properties are rarely reported in phonon systems.
Therefore, one would employ the theory of TQC to find materials with phononic higher-order band topology beyond the electronic materials.

In this work, we first elucidate the physical mechanism that the robust corner modes, which are independent of crystalline symmetries or terminated atoms, can emerge in 2D OAIs.
We explain that such unique feature is classified to the HOTI phase, which can be understood as an intuitive argument how $orbitals$ locate at empty WPs.
Then we adopt the theory of TQC to diagnose phonon band topology and search phononic OAIs in crystalline materials.
Moreover, based on first-principles calculations and RSI theory, we show that the phononic OAIs with robust corner states that are not dependent on the clipping forms of nano-disks can be realized in the $MX_3$ ($M$=Bi, Sb, As, Sc,Y; $X$=I, Br, Cl) monolayers.
Using the TQC theory, we find that the band topology of any odd number of phonon branches in these candidates is not equivalent to a conventional atomic insulator.
Phonon $orbitals$ locating at empty WPs $1a$ and $6i$ are validated, indicating the emergence of HOTI phase with robust corner states.

\begin{figure}[t]
	\includegraphics[width=\linewidth]{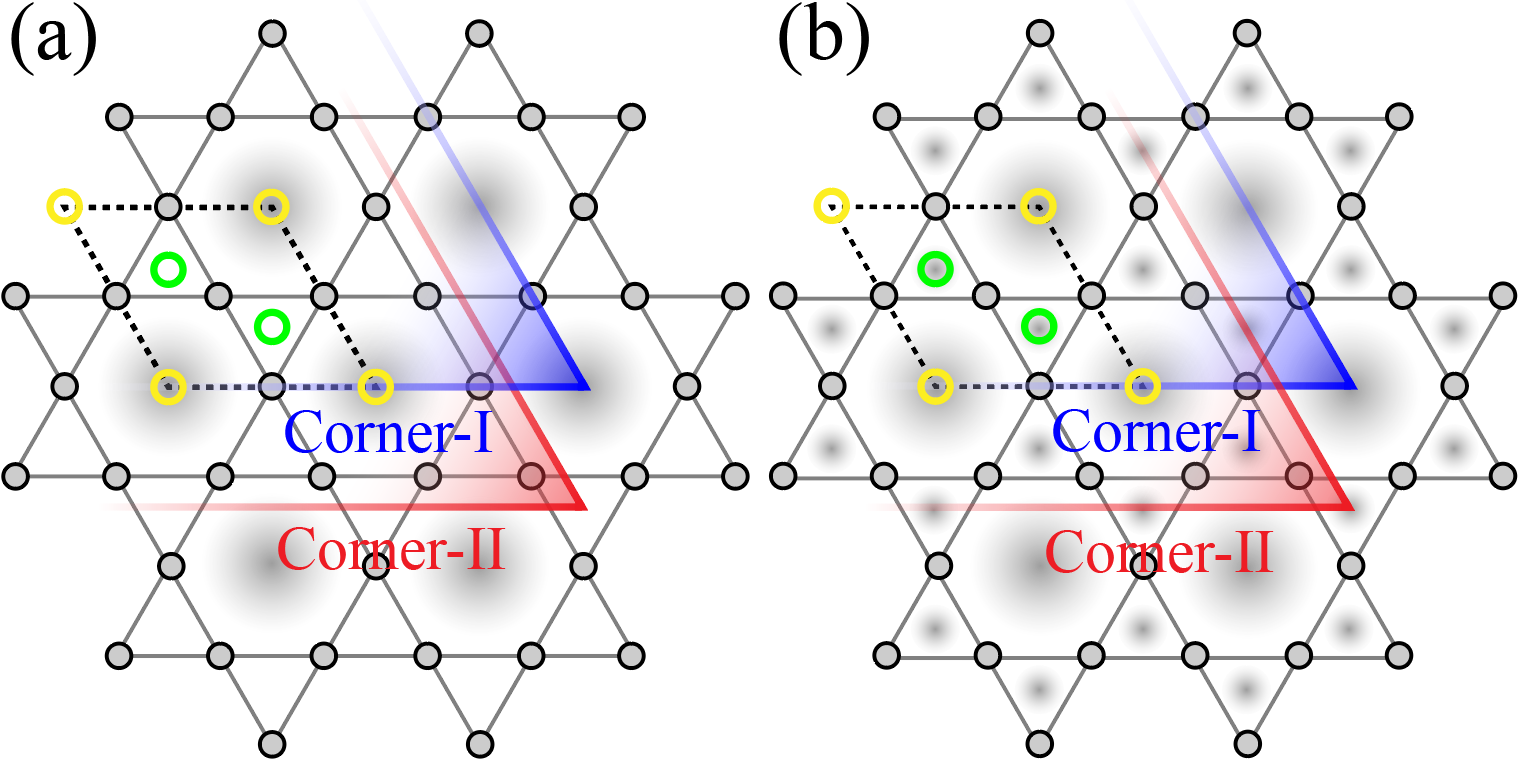}
	\caption{
 (a) The schematic diagram of a kagome-3 lattice in which a set of $orbitals$ \textit{i.e.}, symmetric exponentially decayed Wannier functions are pinned at maximal WP $1a$. The second and third nearest neighbor couplings are not shown. There are two kinds of corners, \textit{i.e.}, Corner-I (in blue) with the OWCC exposed and Corner-II (in red) with the OWCC unexposed.
 (b) Corresponding plots to panel (a) with an additional $orbital$ pinned at $2c$. In panel (b), the OWCC will be exposed in both Corner-I and Corner-II. In both panels, the $1a$ and $2c$ WPs are represented by the circle in yellow and green, respectively.
 }
\label{fig0}
\end{figure}

\textit{OAIs with robust corner modes. ---}
To depict the robust corner modes in HOTIs, we first recall the OAI described by the Su-Schrieffer-Heager (SSH) model, in which edge modes are present when the OWCC is exposed at the edges.
As an extension of SSH model in higher dimension, the corner modes emerge in kagome-3, breathing kagome, and pyrochlore lattices where the band gaps are also in the OAI phase~\cite{PhysRevLett.120.026801,PhysRevLett.126.027002}.
As schematically shown in Fig.~\ref{fig0}, we take the kagome-3 lattice as an example.
There is only one OWCC in this extended SSH model, locating at $1a$.
In this case, when the lattice is clipped into a 0D nano-disk, ingap modes can only be found in the corners with the exposed OWCC, such as the Corner-I highlighted in blue in Fig.~\ref{fig0}(a).
Serving as a contrast to Corner-I, the OWCC is unexposed for Corner-II in Fig.~\ref{fig0}(a), and thus there is no ingap mode located at Corner-II.
As a result, the emergence of the 0D corner states strongly depends on the clipping forms of nano-disks, which brings great challenges to the experimental observation of corner modes.
However, as revealed by the RSI theory \cite{song2020twisted,xu2021three}, the OWCCs for a given OAI is not limited to be single but depend on the realistic decomposition of the BR into EBRs, which means that some orbitals can be allowed to be fixed at two (or more) different empty WPs.
For instance, as shown in Fig.~\ref{fig0}(b), we here assume that an additional $orbital$ acting as an OWCC is pinned at the $2c$ position in kagome-3 lattice.
As a consequence, any kind of corner will expose the OWCC, and thus we can always obtain the corner modes that are robust and independent of crystalline symmetries of clipping forms or cleavage terminations.
In the following, for the first time, we demonstrate that such robust corner modes can be realized in the phonon spectra of $MX_3$ monolayers, forming unique phononic OAIs with attractive HOTI features.

\begin{figure}[b]
	\includegraphics[width=\linewidth]{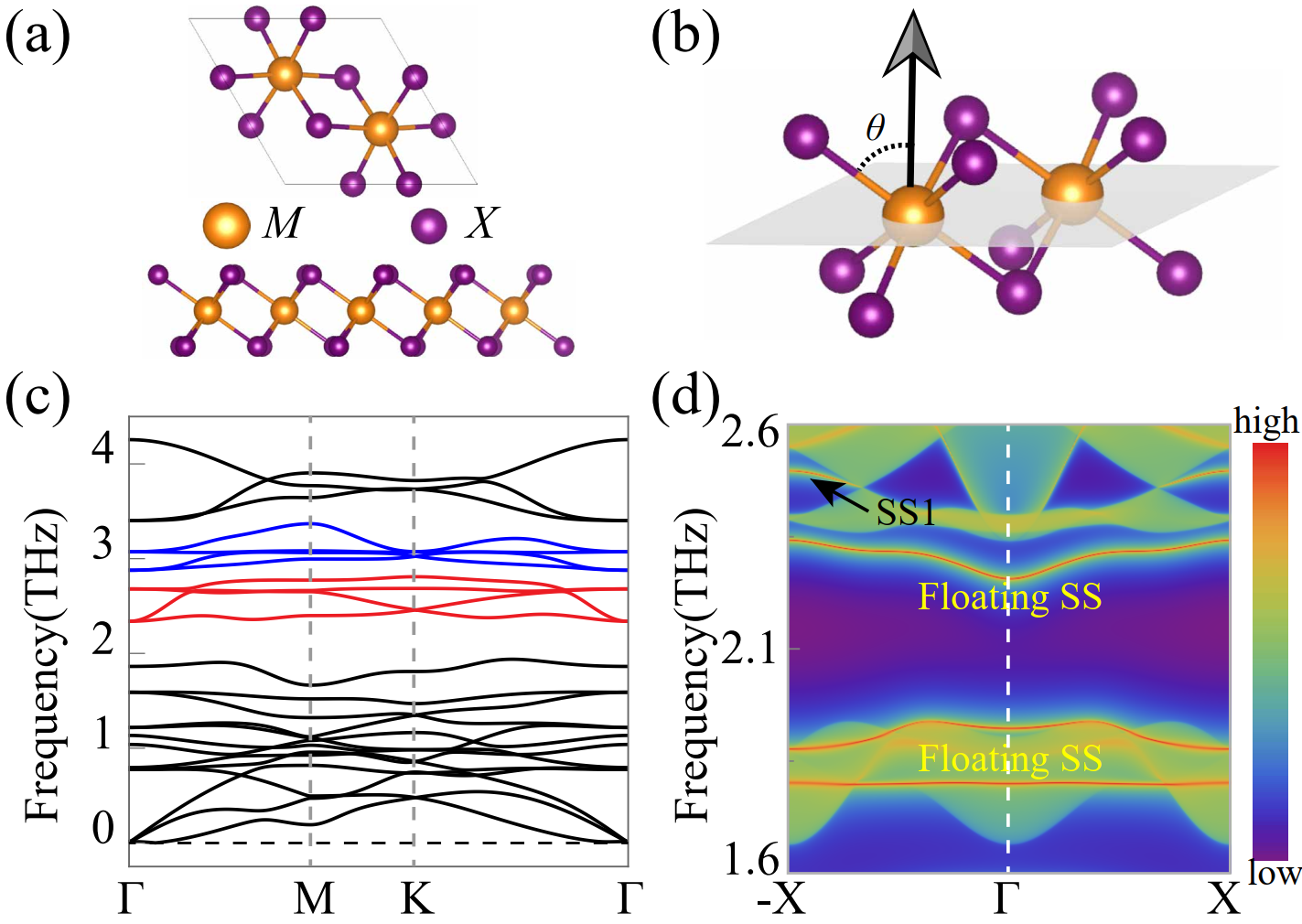}
	\caption{
 (Color online) The structure of $MX_3$ monolayers, and phonon spectrum and surface state of BiI$_3$ monolayer.
 (a) The top and side views of $MX_3$ monolayers.
 (b) The schematic diagram of the definition  of the structure parameter $\theta$, \textit{i.e.}, the angle between the $M$-$X$ bond and the direction normal to the plane.
 (d) The calculated phonon spectrum of BiI$_3$ monolayer. Two special sets of bands formed by $PB_{14}\sim PB_{17}$ and $PB_{18}\sim PB_{21}$ (between 2.0~THz and 3.4~THz,) are highlighted in red and blue, respectively.
 (e) The projected edge states along the zigzag  direction of BiI$_3$ monolayer. The SS1 connect the two projected Dirac points, and floating SSs are observed.
 }
\label{fig1}
\end{figure}

\textit{Lattice structure of $MX_3$ monolayers. ---}
The bulk crystalline structure of $MX_3$ consists of hexagonal $MX_3$ monolayers stacked in ABC configuration, and adjacent monolayers are coupled by van der Waals interactions.
Further researches revealed that the  $MX_3$ monolayers can be exfoliated from their bulk structures~\cite{zhang2016theoretical,liu2017electronic}.
Most recently, this new family of 2D layered semiconductor are synthesised experimentally~\cite{li2017synthesis}.
The $MX_3$ monolayers crystallizes in a hexagonal lattice with space group $P\overline{3}1m$ (No. 162) whose generators are inversion symmetry $P$, three-fold rotational symmetry $C_{3z}$, and mirror symmetry $M_{120}$. 
The WPs $2c ~ \left(1/3,2/3,0\right)$ and $6k ~ \left(x,0,z\right)$ are occupied by the metallic atoms $M$ and halogen atoms $X$, respectively.
As shown in Fig.~\ref{fig1} (a), in $MX_3$ monolayers, each $M$ atom is surrounded by six $X$ atoms that are connected with each other by $P$ and $C_{3z}$.
Here, we take BiI$_3$ as an example and obtain the optimized lattice constant $a=7.83$ {\AA}, the nearest neighbor Bi-I distance $d=3.12$ {\AA}, and the angle between Bi-I bond and the direction normal to the plane $\theta=54.26^{\circ}$, schematically shown in Fig.~\ref{fig1} (b), through  structural optimization.
We tabulate the lattice parameters of $MX_3$ monolayers and experimental values of corresponding bulk structure in Table {\color{blue}S1} in Supplemental Materials (SM)~\cite{SUPP}.
The optimized parameters are in good agreement with the experimental values.

\textit{Phonon spectra of $MX_3$ monolayers. ---}
To obtain phonon spectra of $MX_3$ monolayers, we carried out first-principles calculations as implemented in the Vienna \textit{ab initio} simulation package \cite{PhysRevB.54.11169} and the interatomic force constants were calculated by finite displacements \cite{togo2015first} (see the details in SM~\cite{SUPP}). In the main text, we mainly focus on the BiI$_3$ monolayer and the results of other candidates are included in the SM~\cite{SUPP}. The calculated phonon spectrum of BiI$_3$ monolayer is shown in Fig.~\ref{fig1}(c).
Originating from eight atoms per unit cell, there are 24 phonon branches, labeled as $PB_{k=1,2,\cdots,24}$ ordered by frequency from 0~THz to $\sim4.3$~THz.
Consistent with the characteristic feature of phonon spectrum of 2D materials, there is one transverse acoustic branch exhibiting quadratic dispersion near the $\Gamma$ point.
It is found that all of the 24 branches can be divided into five sets by four visible phonon band gaps.
Moreover, there are two sets of branches between 2.0~THz and 3.4~THz, as highlighted in red and blue in Fig.~\ref{fig1}(c). 
Each set contains the unique quadruple-branch structure along the high-symmetry path $\Gamma$-M-K-$\Gamma$, similar to hourglass fermions \cite{wang2016hourglass}, \textit{i.e.}, two double-degenerated points at $\Gamma$ give rise to a branch-switching pattern along this path.
This typical feature can be obtained by putting $E$ orbital at the $2c$ WP of space group $P\overline{3}1m$.
Taking the set of red phonon branches in Fig.~\ref{fig1}(c) as an example, the quadratic branch dispersion at $\Gamma$ and the linear branch crossing at K point appear near 2.4~THz.
Noteworthy, as an analogy of that in graphene, the band topology of the linear Dirac point at K point implies that there is edge states connecting the projections of two gapless points.
To verify this, we compute the phonon edge spectrum with an open boundary condition along the zigzag direction.
As marked by SS1 in Fig.~\ref{fig1}(d), the expectant edge state is obtained.
Remarkably, in addition to the SS1 originating from the Dirac point, there are three other floating edge states, which is different from the 2D $\mathbb{Z}_2$ TI where the edge states connect the valence and conduction bands.

\begin{table*}
\caption{
The topological properties of $MX_3$ with odd occupied phonon branches, \textit{i.e.}, $N_{Occ}$.
\CheckmarkBold means the set of occupied phonon branches are inequivalent to atomic insulators, and \XSolidBrush means there is no band gap.
Here, we only list the odd $N_{Occ}$s where a band gap can be found in at least one phonon spectrum of $MX_3$ monolayers.
}
\vspace{0.2cm}
\renewcommand\arraystretch{1.8}
\setlength{\tabcolsep}{4.8mm}{
\begin{tabular}{ccccccccccc}
\hline
\hline
$N_{Occ}$   & BiI$_3$ & SbI$_3$ & ScI$_3$ & YI$_3$ & AsI$_3$ & ScBr$_3$ & YBr$_3$ & BiBr$_3$ & YCl$_3$ & ScCl$_3$ \tabularnewline
\hline
 21  & \CheckmarkBold & \CheckmarkBold & \CheckmarkBold & \CheckmarkBold  & \CheckmarkBold &\CheckmarkBold  &\CheckmarkBold  &\XSolidBrush    &\XSolidBrush    &  \CheckmarkBold  \tabularnewline
 19  & \XSolidBrush   & \XSolidBrush   & \CheckmarkBold & \CheckmarkBold  & \XSolidBrush   &\CheckmarkBold  &\CheckmarkBold  &\CheckmarkBold  &\CheckmarkBold  &  \XSolidBrush    \tabularnewline
 17  & \CheckmarkBold & \CheckmarkBold & \XSolidBrush   & \XSolidBrush    & \CheckmarkBold &\XSolidBrush    &\XSolidBrush    &\CheckmarkBold  &\CheckmarkBold  &  \CheckmarkBold  \tabularnewline
 13  & \CheckmarkBold & \CheckmarkBold & \XSolidBrush   & \XSolidBrush    & \CheckmarkBold &\XSolidBrush    &\XSolidBrush    &\CheckmarkBold  &\XSolidBrush    &  \XSolidBrush    \tabularnewline
\hline
\hline
\end{tabular}}
\label{tab1}
\end{table*}

\begin{figure}
	\includegraphics[width=\linewidth]{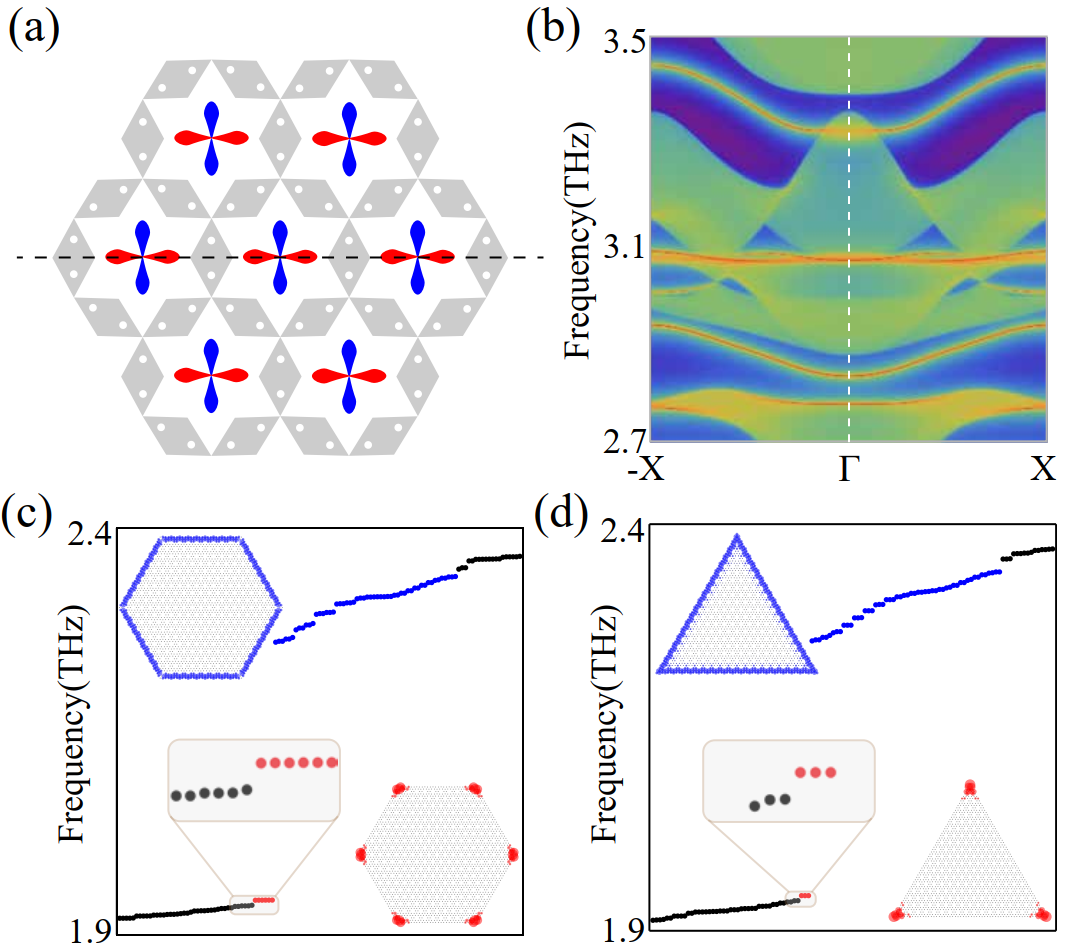}
	\caption{
 (a) The schematic diagram of BiI$_3$ monolayer with a $E_u$ $orbital$ at $1a$. The $6i$ position where a $B$ orbital is located at is represented by the dots in white.  The black dashed line marks the zigzag direction edge with the OWCCs are exposed.
 (b) The projected edge states along the zigzag  direction of BiI$_3$ monolayer.
 (c) The phonon spectrum for a hexagonal-shaped nano-disk. The corner, edge, and bulk states are marked in red, blue, and black, respectively. The spatial distribution of the corner and edge phonon modes are shown as the inserts.
 (d) Corresponding plot to panel (c) starting from the triangular-shaped nano-disk.
 }
\label{fig3}
\end{figure}

\textit{Filling-enforced OAI in $MX_3$ monolayers. ---}
Here, we use the theory of TQC to diagnose the band topology and its novel properties in the phonon spectra of $MX_3$ monolayers.
We calculated the representations of phonon branches at the maximal $k$-vectors (see the SM~\cite{SUPP}).
We find that, for each of the four gaps, the BR can be written as a sum of EBRs, indicating these gaps are in atomic insulator or OAI phases, \textit{i.e.}, are identified as $trivial$ insulators by TQC.
Then, we give a brief introduction of filling-enforced OAI (feOAI) to reveal the phononic OAI phase.
This kind of OAI can be identified just by counting the number of occupied bands without further $ab$-$initio$ calculations.
Without loss of generality, we assume a system belongs to space group $\boldsymbol{G}$, and the WPs ($\omega_{1},\omega_{2},\cdots,\omega_{M}$) are occupied.
The site symmetry of the $\omega_{i}$ WP is $\boldsymbol{g}_{\omega_{i}}$, and the $j$-th irreducible representations (Irrps) of $\boldsymbol{g}_{\omega_{i}}$ are $\rho_{\omega_{i}}^{j=1,2,\cdots,N_{rep,\omega_{i}}}$.
According to the theory of TQC, the Irrps $\rho_{\omega_{i}}^{j}$ in space group $\boldsymbol{G}$ can be expressed as $\ensuremath{\rho_{\omega_{i}}^{j}}\uparrow\boldsymbol{G}$. The corresponding dimension is denoted as $d\left(\ensuremath{\rho_{\omega_{i}}^{j}}\uparrow\boldsymbol{G}\right)$.
In this case, for a $trivial$ insulator the necessary and sufficient conditions for a feOAI is~\cite{xu2021filling}
\begin{eqnarray}
\nexists N_{i,j}\geq0,\in\mathrm{N},s.t.N_{Occ} & = & \sum_{i}^{M}\sum_{j}^{N_{rep,\omega_{i}}}d\left(\ensuremath{\rho_{\omega_{i}}^{j}}\uparrow\boldsymbol{G}\right)N_{i,j}, \label{feOAI}
\end{eqnarray}
with $N_{Occ}$ and $N_{i,j}$ are numbers of occupied bands and $orbitals$ ($\ensuremath{\rho_{\omega_{i}}^{j}}\uparrow\boldsymbol{G}$), respectively. The parameter $j$ sums over all of the occupied WPs, and $N_{i,j}$ are required to be a non-negative integer.

Now, we adopt the concept of feOAI to verify phonon band topology of $MX_3$, where the occupied WPs are $2c$ and $6k$.
The possible dimensions of Irrps that are induced from $orbitals$ at these two WPs are $d\left(\ensuremath{\rho_{2c}^{j}}\uparrow\boldsymbol{G}\right)=2$ or $4$, and $d\left(\ensuremath{\rho_{6k}^{j}}\uparrow\boldsymbol{G}\right)=6$.
Thus, an necessary condition for $MX_3$ monolayers belonged to a conventional atomic insulator is $N_{Occ}=2l$ with non-negative integer $l$.
Equivalently, if a set of phonon branches whose BR is diagnosed as a sum of EBRs and the corresponding $N_{Occ}$ is odd, this set of phonon branches may belong to the OAI. 
We note that the corresponding $N_{Occ}$s for the four visible band gaps in the phonon spectrum of BiI$_3$ monolayer [see Fig.~\ref{fig1}(c)] are $N_{Occ}=12,13,17$, and $21$.
Hence, except for the gap around 1.7~THz ($N_{Occ}=12$), the other three band gaps are in the OAI phase.
This is the reason for the emergence of the floating SS in Fig.~\ref{fig1}(d).
Meanwhile, the floating SSs are also observed when the $N_{Occ}=17$ and $21$, as shown in Fig.~\ref{fig3}(b).
Besides, the emergence of the OAI also needs the presence of band gaps. As shown in Table \ref{tab1}, we list whether the band gaps are present at the odd $N_{Occ}$ in the phonon spectra of all possible $MX_3$ candidates. 
The results show that the OAI phase always occurs in the $MX_3$ monolayers.

To further confirm the OAI phase in the phonon of $MX_3$, 
based on the BR, we can get one $\mathbb{Z}$ type RSI and one $\mathbb{Z}_2$ type RSI with non-zero integers.
The RSIs defined in real space at $1a$ and $6i$ are
\begin{eqnarray}
\delta_{2}\left(a\right)=-m\left(E_{g}\right)+m\left(E_{u}\right)=1,
\label{hamisplit}
\end{eqnarray}
and
\begin{eqnarray}
\eta_{4}\left(i\right) = \begin{array}{ccc}
\left[-m\left(A\right)+m\left(B\right)\right] & \mathrm{mod} & 2\end{array}=1,
\label{hamisplit}
\end{eqnarray}
where the integer $m\left(\rho_{\omega_{i}}^{j}\right)$ means the multiplicy of Irrps $\rho_{\omega_{i}}^{j}$.
The RSI $\delta_{2}\left(a\right)=\eta_{4}\left(i\right)=1$ implies that there is at least one $E_{u}$ and one $B$ $orbital$ pinned at unoccupied WPs
 $1a$ and $6i$, respectively.
Under the site symmetry group $\overline{3}m$ ($2$), for $1a$ ($6i$) position, the IRR $E_{u}$ ($B$) can be induced from $p$ orbitals.
Thus, as an example, when the cleavage termination cuts through the black dashed line in Fig.~\ref{fig3}(a), that is, the $1a$ or $6i$ position is exposed to the edge, the floating SSs can be present [see Fig.~\ref{fig1}(d) and Fig.~\ref{fig3}(b)].
For gaps with $N_{Occ}=17$ and $21$, as shown in the SM~\cite{SUPP}, we also get non-zero RSI, which indicates that these two gaps are also in the OAI phase.

\textit{HOTIs with robust corner states. ---}
It has been reported that the mismatch between the Wannier charge centers and the occupied WPs indicates the existence of higher-order band topology.
As an essential feature, this mismatch is naturally satisfied in OAIs.
Thus, we should have phonon corner modes appeared in gaps of $MX_3$ since the phonon spectra of these monolayers are in the OAI phase.
For instance, we consider the gap between $PB_{13}$ and $PB_{14}$, \textit{i.e.}, that between 1.938~THz and 2.339~THz in the phonon spectrum of BiI$_3$.
To directly show the corner states, we calculate the energy spectrum for a hexagonal-shaped nano-disk of BiI$_3$, that preserves $C_{3z}$, $P$, and $M_{120}$ symmetries with open boundaries. 
As expected, we find that there are six degenerate modes at 1.943~THz and many non-degenerate modes emerge in the gap, as highlighted in red and blue in Fig.~\ref{fig3}(c), respectively. 
The spatial distributions of six degenerate modes are symmetrically localized at the six corners of the nano-disk, indicating that these modes are phonon corner states.
As a comparison, as plotted in blue in Fig.~\ref{fig3} (c), the other modes in the gap are located at the six edges of the nano-disk, and thus they are the edge states. 
As shown in Fig.~\ref{fig3}(d), the similar calculations are done for a triangular-shaped nano-disk where the inversion symmetry $P$ is broken. 
We also obtain three degenerate modes at 1.943~THz corresponding to corner states.
Therefore, we conclude that the phononic HOTI phase is indeed present in the BiI$_3$ monolayer. 
As discussed in SM~\cite{SUPP}, we demonstrate that this HOTI phase beyond the framework of real Chern insulator~\cite{zhao2020equivariant} where corner modes are also reported.  


\begin{figure}[b]
	\includegraphics[width=\linewidth]{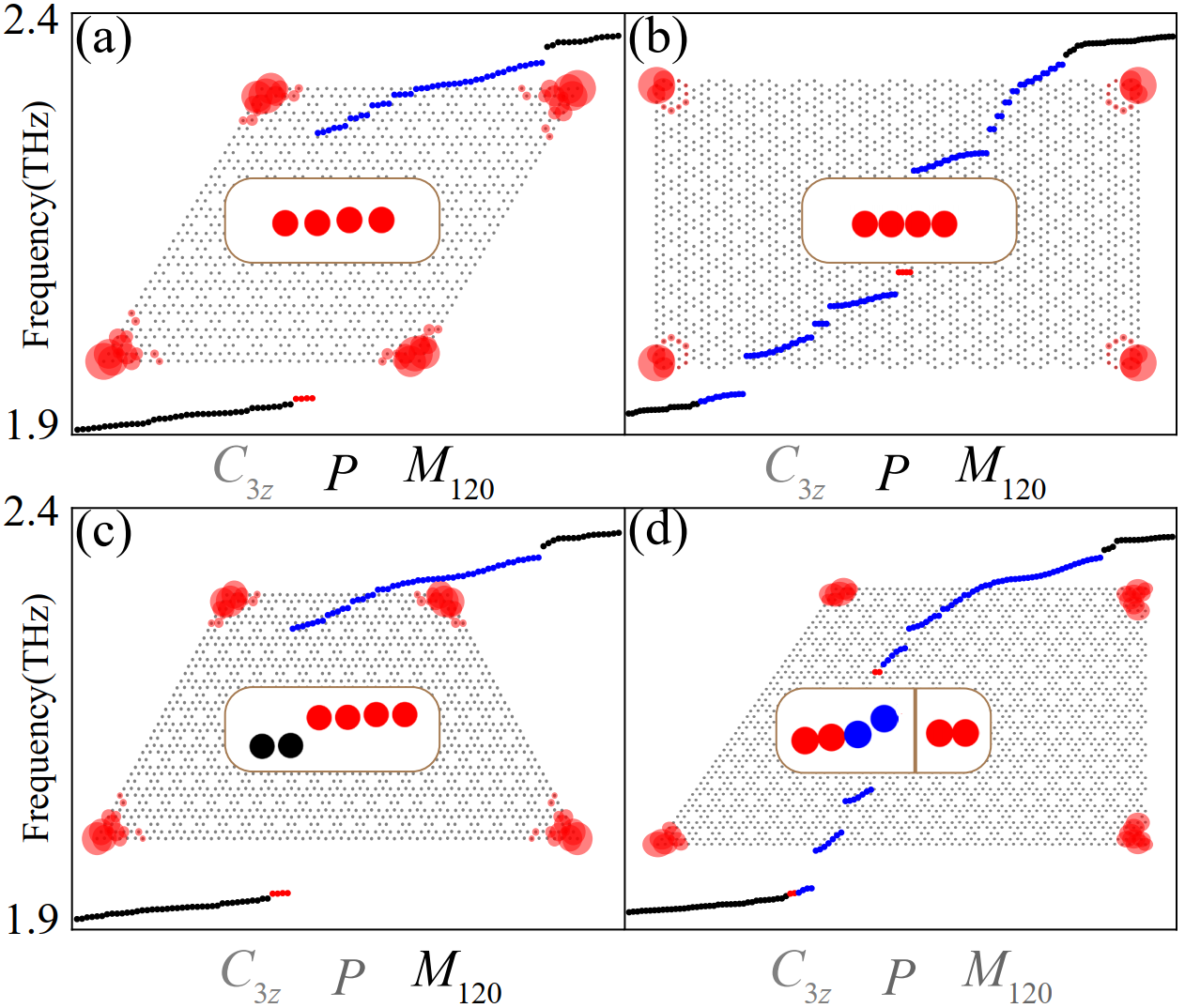}
	\caption{
 (a)-(d) The computed spatial distributions of emerging corner states in the energy spectra for a  (a) rhombic-, (b) rectangular-, (c) trapezoidal-, and (d) semi-trapezoidal-shaped nano-disks. The broken and maintained symmetries are marked in gray and black, respectively.  The corner, edge, and bulk states are marked in red, blue, and black, respectively. The spatial distribution of the corner modes are inserted.
 }
\label{fig4}
\end{figure}

More importantly, since the $orbitals$ pinned at $1a$ or $6i$ will be always been exposed in an arbitrary corner of the $MX_3$ monolayers, the emergence of phonon corner states will be independent of clipping forms or cleavage terminations of nano-disk geometries, which is in sharp contrast to the HOTIs in previous works.
To confirm this prediction, we respectively compute the frequency spectra for a rhombic-, rectangular-, trapezoidal-, and semi-trapezoidal-shaped nano-disks, as shown in Figs.~\ref{fig4}(a)-\ref{fig4}(d). 
The $C_{3z}$ symmetry is broken in the rhombic- and  rectangular-shaped nano-disks, and the $C_{3z}$ and $P$ ($C_{3z}$, $P$, and $M_{120}$) symmetries are broken in the trapezoidal-shaped (semi-trapezoidal-shaped) nano-disks. 
In these four different nano-disks, we can always find phonon corner modes in the gap, and their spatial distributions localized at the corners are clearly visible. 
In the SM~\cite{SUPP}, we also revealed that the robustness of phonon corner states by investigate the energy spectrum of nano-disks with different cleavage terminations of atoms.

\textit{Summary. ---}
In summary, using the theory of TQC, we established a scheme to realize robust corner states in 2D OAIs based an intuitive argument how OWCCs locate at empty WPs.
We emphasize that these corner states are independent of crystalline symmetries or terminated atoms of nano-disk geometries, exhibiting uniquely robust features of HOTI phases.
Then, for the first time, we adopt the theory of TQC to diagnose phonon band topology and search phononic OAIs in 2D crystalline materials.
We further show by first-principles calculations that the phononic OAIs with robust corner states can be realized in the 2D $MX_3$ monolayers.
With the help of feOAI and RSI concepts, we proved that the gaps in the phonon spectra of $MX_3$ with odd occupied states are in the OAI phase.
To verify these promising features of HOTIs, we obtained the phonon corner modes in different shapes or cleavage terminations of nano-disks.
The robustness of corner states would facilitate to explore more attractive phenomena of higher-order band topology not only in phonon systems but also in electronic materials and artificial periodic systems.
Besides, our findings would also advance further applications in low-dimensional devices.

\textit{Acknowledgments. ---}
This work was supported by the National Natural Science Foundation of China (NSFC, Grants No.~12204074, No.~12222402, No.~11974062, and No. 12147102) and the Shenzhen Institute for Quantum Science and Engineering (Grant No. SIQSE202101). 
D.-S. Ma also acknowledges the funding from the China National Postdoctoral Program for Innovative Talent (Grant No. BX20220367).

\bibliography{ref}

\begin{thebibliography}{74}%
\makeatletter
\providecommand \@ifxundefined [1]{%
 \@ifx{#1\undefined}
}%
\providecommand \@ifnum [1]{%
 \ifnum #1\expandafter \@firstoftwo
 \else \expandafter \@secondoftwo
 \fi
}%
\providecommand \@ifx [1]{%
 \ifx #1\expandafter \@firstoftwo
 \else \expandafter \@secondoftwo
 \fi
}%
\providecommand \natexlab [1]{#1}%
\providecommand \enquote  [1]{``#1''}%
\providecommand \bibnamefont  [1]{#1}%
\providecommand \bibfnamefont [1]{#1}%
\providecommand \citenamefont [1]{#1}%
\providecommand \href@noop [0]{\@secondoftwo}%
\providecommand \href [0]{\begingroup \@sanitize@url \@href}%
\providecommand \@href[1]{\@@startlink{#1}\@@href}%
\providecommand \@@href[1]{\endgroup#1\@@endlink}%
\providecommand \@sanitize@url [0]{\catcode `\\12\catcode `\$12\catcode
  `\&12\catcode `\#12\catcode `\^12\catcode `\_12\catcode `\%12\relax}%
\providecommand \@@startlink[1]{}%
\providecommand \@@endlink[0]{}%
\providecommand \url  [0]{\begingroup\@sanitize@url \@url }%
\providecommand \@url [1]{\endgroup\@href {#1}{\urlprefix }}%
\providecommand \urlprefix  [0]{URL }%
\providecommand \Eprint [0]{\href }%
\providecommand \doibase [0]{http://dx.doi.org/}%
\providecommand \selectlanguage [0]{\@gobble}%
\providecommand \bibinfo  [0]{\@secondoftwo}%
\providecommand \bibfield  [0]{\@secondoftwo}%
\providecommand \translation [1]{[#1]}%
\providecommand \BibitemOpen [0]{}%
\providecommand \bibitemStop [0]{}%
\providecommand \bibitemNoStop [0]{.\EOS\space}%
\providecommand \EOS [0]{\spacefactor3000\relax}%
\providecommand \BibitemShut  [1]{\csname bibitem#1\endcsname}%
\let\auto@bib@innerbib\@empty
\bibitem [{\citenamefont {Benalcazar}\ \emph {et~al.}(2017)\citenamefont
  {Benalcazar}, \citenamefont {Bernevig},\ and\ \citenamefont
  {Hughes}}]{benalcazar2017quantized}%
  \BibitemOpen
  \bibfield  {author} {\bibinfo {author} {\bibfnamefont {W.~A.}\ \bibnamefont
  {Benalcazar}}, \bibinfo {author} {\bibfnamefont {B.~A.}\ \bibnamefont
  {Bernevig}}, \ and\ \bibinfo {author} {\bibfnamefont {T.~L.}\ \bibnamefont
  {Hughes}},\ }\href {https://www.science.org/doi/10.1126/science.aah6442}
  {\bibfield  {journal} {\bibinfo  {journal} {Science}\ }\textbf {\bibinfo
  {volume} {357}},\ \bibinfo {pages} {61} (\bibinfo {year} {2017})}\BibitemShut
  {NoStop}%
\bibitem [{\citenamefont {Peterson}\ \emph {et~al.}(2018)\citenamefont
  {Peterson}, \citenamefont {Benalcazar}, \citenamefont {Hughes},\ and\
  \citenamefont {Bahl}}]{peterson2018quantized}%
  \BibitemOpen
  \bibfield  {author} {\bibinfo {author} {\bibfnamefont {C.~W.}\ \bibnamefont
  {Peterson}}, \bibinfo {author} {\bibfnamefont {W.~A.}\ \bibnamefont
  {Benalcazar}}, \bibinfo {author} {\bibfnamefont {T.~L.}\ \bibnamefont
  {Hughes}}, \ and\ \bibinfo {author} {\bibfnamefont {G.}~\bibnamefont
  {Bahl}},\ }\href {https://www.nature.com/articles/nature25777} {\bibfield
  {journal} {\bibinfo  {journal} {Nature}\ }\textbf {\bibinfo {volume} {555}},\
  \bibinfo {pages} {346} (\bibinfo {year} {2018})}\BibitemShut {NoStop}%
\bibitem [{\citenamefont {Schindler}\ \emph {et~al.}(2018)\citenamefont
  {Schindler}, \citenamefont {Cook}, \citenamefont {Vergniory}, \citenamefont
  {Wang}, \citenamefont {Parkin}, \citenamefont {Bernevig},\ and\ \citenamefont
  {Neupert}}]{schindler2018higher}%
  \BibitemOpen
  \bibfield  {author} {\bibinfo {author} {\bibfnamefont {F.}~\bibnamefont
  {Schindler}}, \bibinfo {author} {\bibfnamefont {A.~M.}\ \bibnamefont {Cook}},
  \bibinfo {author} {\bibfnamefont {M.~G.}\ \bibnamefont {Vergniory}}, \bibinfo
  {author} {\bibfnamefont {Z.}~\bibnamefont {Wang}}, \bibinfo {author}
  {\bibfnamefont {S.~S.}\ \bibnamefont {Parkin}}, \bibinfo {author}
  {\bibfnamefont {B.~A.}\ \bibnamefont {Bernevig}}, \ and\ \bibinfo {author}
  {\bibfnamefont {T.}~\bibnamefont {Neupert}},\ }\href
  {https://www.science.org/doi/10.1126/sciadv.aat0346} {\bibfield  {journal}
  {\bibinfo  {journal} {Sci. Adv.}\ }\textbf {\bibinfo {volume} {4}},\ \bibinfo
  {pages} {eaat0346} (\bibinfo {year} {2018})}\BibitemShut {NoStop}%
\bibitem [{\citenamefont {Song}\ \emph {et~al.}(2017)\citenamefont {Song},
  \citenamefont {Fang},\ and\ \citenamefont {Fang}}]{PhysRevLett.119.246402}%
  \BibitemOpen
  \bibfield  {author} {\bibinfo {author} {\bibfnamefont {Z.}~\bibnamefont
  {Song}}, \bibinfo {author} {\bibfnamefont {Z.}~\bibnamefont {Fang}}, \ and\
  \bibinfo {author} {\bibfnamefont {C.}~\bibnamefont {Fang}},\ }\href {\doibase
  10.1103/PhysRevLett.119.246402} {\bibfield  {journal} {\bibinfo  {journal}
  {Phys. Rev. Lett.}\ }\textbf {\bibinfo {volume} {119}},\ \bibinfo {pages}
  {246402} (\bibinfo {year} {2017})}\BibitemShut {NoStop}%
\bibitem [{\citenamefont {Langbehn}\ \emph {et~al.}(2017)\citenamefont
  {Langbehn}, \citenamefont {Peng}, \citenamefont {Trifunovic}, \citenamefont
  {von Oppen},\ and\ \citenamefont {Brouwer}}]{PhysRevLett.119.246401}%
  \BibitemOpen
  \bibfield  {author} {\bibinfo {author} {\bibfnamefont {J.}~\bibnamefont
  {Langbehn}}, \bibinfo {author} {\bibfnamefont {Y.}~\bibnamefont {Peng}},
  \bibinfo {author} {\bibfnamefont {L.}~\bibnamefont {Trifunovic}}, \bibinfo
  {author} {\bibfnamefont {F.}~\bibnamefont {von Oppen}}, \ and\ \bibinfo
  {author} {\bibfnamefont {P.~W.}\ \bibnamefont {Brouwer}},\ }\href {\doibase
  10.1103/PhysRevLett.119.246401} {\bibfield  {journal} {\bibinfo  {journal}
  {Phys. Rev. Lett.}\ }\textbf {\bibinfo {volume} {119}},\ \bibinfo {pages}
  {246401} (\bibinfo {year} {2017})}\BibitemShut {NoStop}%
\bibitem [{\citenamefont {Khalaf}(2018)}]{PhysRevB.97.205136}%
  \BibitemOpen
  \bibfield  {author} {\bibinfo {author} {\bibfnamefont {E.}~\bibnamefont
  {Khalaf}},\ }\href {\doibase 10.1103/PhysRevB.97.205136} {\bibfield
  {journal} {\bibinfo  {journal} {Phys. Rev. B}\ }\textbf {\bibinfo {volume}
  {97}},\ \bibinfo {pages} {205136} (\bibinfo {year} {2018})}\BibitemShut
  {NoStop}%
\bibitem [{\citenamefont {Wang}\ \emph {et~al.}(2019)\citenamefont {Wang},
  \citenamefont {Wieder}, \citenamefont {Li}, \citenamefont {Yan},\ and\
  \citenamefont {Bernevig}}]{PhysRevLett.123.186401}%
  \BibitemOpen
  \bibfield  {author} {\bibinfo {author} {\bibfnamefont {Z.}~\bibnamefont
  {Wang}}, \bibinfo {author} {\bibfnamefont {B.~J.}\ \bibnamefont {Wieder}},
  \bibinfo {author} {\bibfnamefont {J.}~\bibnamefont {Li}}, \bibinfo {author}
  {\bibfnamefont {B.}~\bibnamefont {Yan}}, \ and\ \bibinfo {author}
  {\bibfnamefont {B.~A.}\ \bibnamefont {Bernevig}},\ }\href {\doibase
  10.1103/PhysRevLett.123.186401} {\bibfield  {journal} {\bibinfo  {journal}
  {Phys. Rev. Lett.}\ }\textbf {\bibinfo {volume} {123}},\ \bibinfo {pages}
  {186401} (\bibinfo {year} {2019})}\BibitemShut {NoStop}%
\bibitem [{\citenamefont {Benalcazar}\ \emph {et~al.}(2019)\citenamefont
  {Benalcazar}, \citenamefont {Li},\ and\ \citenamefont
  {Hughes}}]{PhysRevB.99.245151}%
  \BibitemOpen
  \bibfield  {author} {\bibinfo {author} {\bibfnamefont {W.~A.}\ \bibnamefont
  {Benalcazar}}, \bibinfo {author} {\bibfnamefont {T.}~\bibnamefont {Li}}, \
  and\ \bibinfo {author} {\bibfnamefont {T.~L.}\ \bibnamefont {Hughes}},\
  }\href {\doibase 10.1103/PhysRevB.99.245151} {\bibfield  {journal} {\bibinfo
  {journal} {Phys. Rev. B}\ }\textbf {\bibinfo {volume} {99}},\ \bibinfo
  {pages} {245151} (\bibinfo {year} {2019})}\BibitemShut {NoStop}%
\bibitem [{\citenamefont {Xu}\ \emph {et~al.}(2019)\citenamefont {Xu},
  \citenamefont {Song}, \citenamefont {Wang}, \citenamefont {Weng},\ and\
  \citenamefont {Dai}}]{PhysRevLett.122.256402}%
  \BibitemOpen
  \bibfield  {author} {\bibinfo {author} {\bibfnamefont {Y.}~\bibnamefont
  {Xu}}, \bibinfo {author} {\bibfnamefont {Z.}~\bibnamefont {Song}}, \bibinfo
  {author} {\bibfnamefont {Z.}~\bibnamefont {Wang}}, \bibinfo {author}
  {\bibfnamefont {H.}~\bibnamefont {Weng}}, \ and\ \bibinfo {author}
  {\bibfnamefont {X.}~\bibnamefont {Dai}},\ }\href {\doibase
  10.1103/PhysRevLett.122.256402} {\bibfield  {journal} {\bibinfo  {journal}
  {Phys. Rev. Lett.}\ }\textbf {\bibinfo {volume} {122}},\ \bibinfo {pages}
  {256402} (\bibinfo {year} {2019})}\BibitemShut {NoStop}%
\bibitem [{\citenamefont {Sheng}\ \emph {et~al.}(2019)\citenamefont {Sheng},
  \citenamefont {Chen}, \citenamefont {Liu}, \citenamefont {Chen},
  \citenamefont {Yu}, \citenamefont {Zhao},\ and\ \citenamefont
  {Yang}}]{PhysRevLett.123.256402}%
  \BibitemOpen
  \bibfield  {author} {\bibinfo {author} {\bibfnamefont {X.-L.}\ \bibnamefont
  {Sheng}}, \bibinfo {author} {\bibfnamefont {C.}~\bibnamefont {Chen}},
  \bibinfo {author} {\bibfnamefont {H.}~\bibnamefont {Liu}}, \bibinfo {author}
  {\bibfnamefont {Z.}~\bibnamefont {Chen}}, \bibinfo {author} {\bibfnamefont
  {Z.-M.}\ \bibnamefont {Yu}}, \bibinfo {author} {\bibfnamefont {Y.~X.}\
  \bibnamefont {Zhao}}, \ and\ \bibinfo {author} {\bibfnamefont {S.~A.}\
  \bibnamefont {Yang}},\ }\href {\doibase 10.1103/PhysRevLett.123.256402}
  {\bibfield  {journal} {\bibinfo  {journal} {Phys. Rev. Lett.}\ }\textbf
  {\bibinfo {volume} {123}},\ \bibinfo {pages} {256402} (\bibinfo {year}
  {2019})}\BibitemShut {NoStop}%
\bibitem [{\citenamefont {Zhang}\ \emph {et~al.}(2020)\citenamefont {Zhang},
  \citenamefont {Wu},\ and\ \citenamefont
  {Das~Sarma}}]{PhysRevLett.124.136407}%
  \BibitemOpen
  \bibfield  {author} {\bibinfo {author} {\bibfnamefont {R.-X.}\ \bibnamefont
  {Zhang}}, \bibinfo {author} {\bibfnamefont {F.}~\bibnamefont {Wu}}, \ and\
  \bibinfo {author} {\bibfnamefont {S.}~\bibnamefont {Das~Sarma}},\ }\href
  {\doibase 10.1103/PhysRevLett.124.136407} {\bibfield  {journal} {\bibinfo
  {journal} {Phys. Rev. Lett.}\ }\textbf {\bibinfo {volume} {124}},\ \bibinfo
  {pages} {136407} (\bibinfo {year} {2020})}\BibitemShut {NoStop}%
\bibitem [{\citenamefont {Chen}\ \emph {et~al.}(2020)\citenamefont {Chen},
  \citenamefont {Song}, \citenamefont {Zhao}, \citenamefont {Chen},
  \citenamefont {Yu}, \citenamefont {Sheng},\ and\ \citenamefont
  {Yang}}]{PhysRevLett.125.056402}%
  \BibitemOpen
  \bibfield  {author} {\bibinfo {author} {\bibfnamefont {C.}~\bibnamefont
  {Chen}}, \bibinfo {author} {\bibfnamefont {Z.}~\bibnamefont {Song}}, \bibinfo
  {author} {\bibfnamefont {J.-Z.}\ \bibnamefont {Zhao}}, \bibinfo {author}
  {\bibfnamefont {Z.}~\bibnamefont {Chen}}, \bibinfo {author} {\bibfnamefont
  {Z.-M.}\ \bibnamefont {Yu}}, \bibinfo {author} {\bibfnamefont {X.-L.}\
  \bibnamefont {Sheng}}, \ and\ \bibinfo {author} {\bibfnamefont {S.~A.}\
  \bibnamefont {Yang}},\ }\href {\doibase 10.1103/PhysRevLett.125.056402}
  {\bibfield  {journal} {\bibinfo  {journal} {Phys. Rev. Lett.}\ }\textbf
  {\bibinfo {volume} {125}},\ \bibinfo {pages} {056402} (\bibinfo {year}
  {2020})}\BibitemShut {NoStop}%
\bibitem [{\citenamefont {Fan}\ \emph {et~al.}(2019)\citenamefont {Fan},
  \citenamefont {Xia}, \citenamefont {Tong}, \citenamefont {Zheng},\ and\
  \citenamefont {Yu}}]{PhysRevLett.122.204301}%
  \BibitemOpen
  \bibfield  {author} {\bibinfo {author} {\bibfnamefont {H.}~\bibnamefont
  {Fan}}, \bibinfo {author} {\bibfnamefont {B.}~\bibnamefont {Xia}}, \bibinfo
  {author} {\bibfnamefont {L.}~\bibnamefont {Tong}}, \bibinfo {author}
  {\bibfnamefont {S.}~\bibnamefont {Zheng}}, \ and\ \bibinfo {author}
  {\bibfnamefont {D.}~\bibnamefont {Yu}},\ }\href {\doibase
  10.1103/PhysRevLett.122.204301} {\bibfield  {journal} {\bibinfo  {journal}
  {Phys. Rev. Lett.}\ }\textbf {\bibinfo {volume} {122}},\ \bibinfo {pages}
  {204301} (\bibinfo {year} {2019})}\BibitemShut {NoStop}%
\bibitem [{\citenamefont {Chen}\ \emph {et~al.}(2019)\citenamefont {Chen},
  \citenamefont {Deng}, \citenamefont {Shi}, \citenamefont {Zhao},
  \citenamefont {Chen},\ and\ \citenamefont {Dong}}]{PhysRevLett.122.233902}%
  \BibitemOpen
  \bibfield  {author} {\bibinfo {author} {\bibfnamefont {X.-D.}\ \bibnamefont
  {Chen}}, \bibinfo {author} {\bibfnamefont {W.-M.}\ \bibnamefont {Deng}},
  \bibinfo {author} {\bibfnamefont {F.-L.}\ \bibnamefont {Shi}}, \bibinfo
  {author} {\bibfnamefont {F.-L.}\ \bibnamefont {Zhao}}, \bibinfo {author}
  {\bibfnamefont {M.}~\bibnamefont {Chen}}, \ and\ \bibinfo {author}
  {\bibfnamefont {J.-W.}\ \bibnamefont {Dong}},\ }\href {\doibase
  10.1103/PhysRevLett.122.233902} {\bibfield  {journal} {\bibinfo  {journal}
  {Phys. Rev. Lett.}\ }\textbf {\bibinfo {volume} {122}},\ \bibinfo {pages}
  {233902} (\bibinfo {year} {2019})}\BibitemShut {NoStop}%
\bibitem [{\citenamefont {Lee}\ \emph {et~al.}(2020)\citenamefont {Lee},
  \citenamefont {Kim}, \citenamefont {Ahn},\ and\ \citenamefont
  {Yang}}]{lee2020two}%
  \BibitemOpen
  \bibfield  {author} {\bibinfo {author} {\bibfnamefont {E.}~\bibnamefont
  {Lee}}, \bibinfo {author} {\bibfnamefont {R.}~\bibnamefont {Kim}}, \bibinfo
  {author} {\bibfnamefont {J.}~\bibnamefont {Ahn}}, \ and\ \bibinfo {author}
  {\bibfnamefont {B.-J.}\ \bibnamefont {Yang}},\ }\href
  {https://www.nature.com/articles/s41535-019-0206-8} {\bibfield  {journal}
  {\bibinfo  {journal} {npj Quantum Mater.}\ }\textbf {\bibinfo {volume} {5}},\
  \bibinfo {pages} {1} (\bibinfo {year} {2020})}\BibitemShut {NoStop}%
\bibitem [{\citenamefont {Qian}\ \emph {et~al.}(2021)\citenamefont {Qian},
  \citenamefont {Liu},\ and\ \citenamefont {Yao}}]{PhysRevB.104.245427}%
  \BibitemOpen
  \bibfield  {author} {\bibinfo {author} {\bibfnamefont {S.}~\bibnamefont
  {Qian}}, \bibinfo {author} {\bibfnamefont {C.-C.}\ \bibnamefont {Liu}}, \
  and\ \bibinfo {author} {\bibfnamefont {Y.}~\bibnamefont {Yao}},\ }\href
  {\doibase 10.1103/PhysRevB.104.245427} {\bibfield  {journal} {\bibinfo
  {journal} {Phys. Rev. B}\ }\textbf {\bibinfo {volume} {104}},\ \bibinfo
  {pages} {245427} (\bibinfo {year} {2021})}\BibitemShut {NoStop}%
\bibitem [{\citenamefont {Mu}\ \emph {et~al.}(2022)\citenamefont {Mu},
  \citenamefont {Liu}, \citenamefont {Hu},\ and\ \citenamefont
  {Wang}}]{mu2022kekule}%
  \BibitemOpen
  \bibfield  {author} {\bibinfo {author} {\bibfnamefont {H.}~\bibnamefont
  {Mu}}, \bibinfo {author} {\bibfnamefont {B.}~\bibnamefont {Liu}}, \bibinfo
  {author} {\bibfnamefont {T.}~\bibnamefont {Hu}}, \ and\ \bibinfo {author}
  {\bibfnamefont {Z.}~\bibnamefont {Wang}},\ }\href
  {https://pubs.acs.org/doi/abs/10.1021/acs.nanolett.1c04239} {\bibfield
  {journal} {\bibinfo  {journal} {Nano Lett.}\ }\textbf {\bibinfo {volume}
  {22}},\ \bibinfo {pages} {1122} (\bibinfo {year} {2022})}\BibitemShut
  {NoStop}%
\bibitem [{\citenamefont {Gladstein~Gladstone}\ \emph
  {et~al.}(2022)\citenamefont {Gladstein~Gladstone}, \citenamefont {Jung},\
  and\ \citenamefont {Shvets}}]{PhysRevLett.128.026801}%
  \BibitemOpen
  \bibfield  {author} {\bibinfo {author} {\bibfnamefont {R.}~\bibnamefont
  {Gladstein~Gladstone}}, \bibinfo {author} {\bibfnamefont {M.}~\bibnamefont
  {Jung}}, \ and\ \bibinfo {author} {\bibfnamefont {G.}~\bibnamefont
  {Shvets}},\ }\href {\doibase 10.1103/PhysRevLett.128.026801} {\bibfield
  {journal} {\bibinfo  {journal} {Phys. Rev. Lett.}\ }\textbf {\bibinfo
  {volume} {128}},\ \bibinfo {pages} {026801} (\bibinfo {year}
  {2022})}\BibitemShut {NoStop}%
\bibitem [{\citenamefont {Hasan}\ and\ \citenamefont
  {Kane}(2010)}]{RevModPhys.82.3045}%
  \BibitemOpen
  \bibfield  {author} {\bibinfo {author} {\bibfnamefont {M.~Z.}\ \bibnamefont
  {Hasan}}\ and\ \bibinfo {author} {\bibfnamefont {C.~L.}\ \bibnamefont
  {Kane}},\ }\href {\doibase 10.1103/RevModPhys.82.3045} {\bibfield  {journal}
  {\bibinfo  {journal} {Rev. Mod. Phys.}\ }\textbf {\bibinfo {volume} {82}},\
  \bibinfo {pages} {3045} (\bibinfo {year} {2010})}\BibitemShut {NoStop}%
\bibitem [{\citenamefont {Ezawa}(2018{\natexlab{a}})}]{PhysRevLett.120.026801}%
  \BibitemOpen
  \bibfield  {author} {\bibinfo {author} {\bibfnamefont {M.}~\bibnamefont
  {Ezawa}},\ }\href {\doibase 10.1103/PhysRevLett.120.026801} {\bibfield
  {journal} {\bibinfo  {journal} {Phys. Rev. Lett.}\ }\textbf {\bibinfo
  {volume} {120}},\ \bibinfo {pages} {026801} (\bibinfo {year}
  {2018}{\natexlab{a}})}\BibitemShut {NoStop}%
\bibitem [{\citenamefont {Ezawa}(2018{\natexlab{b}})}]{PhysRevLett.121.116801}%
  \BibitemOpen
  \bibfield  {author} {\bibinfo {author} {\bibfnamefont {M.}~\bibnamefont
  {Ezawa}},\ }\href {\doibase 10.1103/PhysRevLett.121.116801} {\bibfield
  {journal} {\bibinfo  {journal} {Phys. Rev. Lett.}\ }\textbf {\bibinfo
  {volume} {121}},\ \bibinfo {pages} {116801} (\bibinfo {year}
  {2018}{\natexlab{b}})}\BibitemShut {NoStop}%
\bibitem [{\citenamefont {Takahashi}\ \emph {et~al.}(2021)\citenamefont
  {Takahashi}, \citenamefont {Zhang},\ and\ \citenamefont
  {Murakami}}]{PhysRevB.103.205123}%
  \BibitemOpen
  \bibfield  {author} {\bibinfo {author} {\bibfnamefont {R.}~\bibnamefont
  {Takahashi}}, \bibinfo {author} {\bibfnamefont {T.}~\bibnamefont {Zhang}}, \
  and\ \bibinfo {author} {\bibfnamefont {S.}~\bibnamefont {Murakami}},\ }\href
  {\doibase 10.1103/PhysRevB.103.205123} {\bibfield  {journal} {\bibinfo
  {journal} {Phys. Rev. B}\ }\textbf {\bibinfo {volume} {103}},\ \bibinfo
  {pages} {205123} (\bibinfo {year} {2021})}\BibitemShut {NoStop}%
\bibitem [{\citenamefont {Wang}\ \emph {et~al.}(2022)\citenamefont {Wang},
  \citenamefont {Jiang}, \citenamefont {Liu}, \citenamefont {Zhang},
  \citenamefont {Li}, \citenamefont {Liu}, \citenamefont {Sun}, \citenamefont
  {Weng},\ and\ \citenamefont {Chen}}]{wang2022two}%
  \BibitemOpen
  \bibfield  {author} {\bibinfo {author} {\bibfnamefont {L.}~\bibnamefont
  {Wang}}, \bibinfo {author} {\bibfnamefont {Y.}~\bibnamefont {Jiang}},
  \bibinfo {author} {\bibfnamefont {J.}~\bibnamefont {Liu}}, \bibinfo {author}
  {\bibfnamefont {S.}~\bibnamefont {Zhang}}, \bibinfo {author} {\bibfnamefont
  {J.}~\bibnamefont {Li}}, \bibinfo {author} {\bibfnamefont {P.}~\bibnamefont
  {Liu}}, \bibinfo {author} {\bibfnamefont {Y.}~\bibnamefont {Sun}}, \bibinfo
  {author} {\bibfnamefont {H.}~\bibnamefont {Weng}}, \ and\ \bibinfo {author}
  {\bibfnamefont {X.-Q.}\ \bibnamefont {Chen}},\ }\href
  {https://arxiv.org/abs/2205.01994} {\bibfield  {journal} {\bibinfo  {journal}
  {arXiv:2205.01994}\ } (\bibinfo {year} {2022})}\BibitemShut {NoStop}%
\bibitem [{\citenamefont {Bradlyn}\ \emph {et~al.}(2017)\citenamefont
  {Bradlyn}, \citenamefont {Elcoro}, \citenamefont {Cano}, \citenamefont
  {Vergniory}, \citenamefont {Wang}, \citenamefont {Felser}, \citenamefont
  {Aroyo},\ and\ \citenamefont {Bernevig}}]{bradlyn2017topological}%
  \BibitemOpen
  \bibfield  {author} {\bibinfo {author} {\bibfnamefont {B.}~\bibnamefont
  {Bradlyn}}, \bibinfo {author} {\bibfnamefont {L.}~\bibnamefont {Elcoro}},
  \bibinfo {author} {\bibfnamefont {J.}~\bibnamefont {Cano}}, \bibinfo {author}
  {\bibfnamefont {M.}~\bibnamefont {Vergniory}}, \bibinfo {author}
  {\bibfnamefont {Z.}~\bibnamefont {Wang}}, \bibinfo {author} {\bibfnamefont
  {C.}~\bibnamefont {Felser}}, \bibinfo {author} {\bibfnamefont
  {M.}~\bibnamefont {Aroyo}}, \ and\ \bibinfo {author} {\bibfnamefont {B.~A.}\
  \bibnamefont {Bernevig}},\ }\href
  {https://www.nature.com/articles/nature23268} {\bibfield  {journal} {\bibinfo
   {journal} {Nature}\ }\textbf {\bibinfo {volume} {547}},\ \bibinfo {pages}
  {298} (\bibinfo {year} {2017})}\BibitemShut {NoStop}%
\bibitem [{\citenamefont {Cano}\ \emph {et~al.}(2018)\citenamefont {Cano},
  \citenamefont {Bradlyn}, \citenamefont {Wang}, \citenamefont {Elcoro},
  \citenamefont {Vergniory}, \citenamefont {Felser}, \citenamefont {Aroyo},\
  and\ \citenamefont {Bernevig}}]{PhysRevB.97.035139}%
  \BibitemOpen
  \bibfield  {author} {\bibinfo {author} {\bibfnamefont {J.}~\bibnamefont
  {Cano}}, \bibinfo {author} {\bibfnamefont {B.}~\bibnamefont {Bradlyn}},
  \bibinfo {author} {\bibfnamefont {Z.}~\bibnamefont {Wang}}, \bibinfo {author}
  {\bibfnamefont {L.}~\bibnamefont {Elcoro}}, \bibinfo {author} {\bibfnamefont
  {M.~G.}\ \bibnamefont {Vergniory}}, \bibinfo {author} {\bibfnamefont
  {C.}~\bibnamefont {Felser}}, \bibinfo {author} {\bibfnamefont {M.~I.}\
  \bibnamefont {Aroyo}}, \ and\ \bibinfo {author} {\bibfnamefont {B.~A.}\
  \bibnamefont {Bernevig}},\ }\href {\doibase 10.1103/PhysRevB.97.035139}
  {\bibfield  {journal} {\bibinfo  {journal} {Phys. Rev. B}\ }\textbf {\bibinfo
  {volume} {97}},\ \bibinfo {pages} {035139} (\bibinfo {year}
  {2018})}\BibitemShut {NoStop}%
\bibitem [{\citenamefont {Song}\ \emph {et~al.}(2020)\citenamefont {Song},
  \citenamefont {Elcoro},\ and\ \citenamefont {Bernevig}}]{song2020twisted}%
  \BibitemOpen
  \bibfield  {author} {\bibinfo {author} {\bibfnamefont {Z.-D.}\ \bibnamefont
  {Song}}, \bibinfo {author} {\bibfnamefont {L.}~\bibnamefont {Elcoro}}, \ and\
  \bibinfo {author} {\bibfnamefont {B.~A.}\ \bibnamefont {Bernevig}},\ }\href
  {https://www.science.org/doi/abs/10.1126/science.aaz7650} {\bibfield
  {journal} {\bibinfo  {journal} {Science}\ }\textbf {\bibinfo {volume}
  {367}},\ \bibinfo {pages} {794} (\bibinfo {year} {2020})}\BibitemShut
  {NoStop}%
\bibitem [{\citenamefont {Xu}\ \emph {et~al.}(2021{\natexlab{a}})\citenamefont
  {Xu}, \citenamefont {Elcoro}, \citenamefont {Li}, \citenamefont {Song},
  \citenamefont {Regnault}, \citenamefont {Yang}, \citenamefont {Sun},
  \citenamefont {Parkin}, \citenamefont {Felser},\ and\ \citenamefont
  {Bernevig}}]{xu2021three}%
  \BibitemOpen
  \bibfield  {author} {\bibinfo {author} {\bibfnamefont {Y.}~\bibnamefont
  {Xu}}, \bibinfo {author} {\bibfnamefont {L.}~\bibnamefont {Elcoro}}, \bibinfo
  {author} {\bibfnamefont {G.}~\bibnamefont {Li}}, \bibinfo {author}
  {\bibfnamefont {Z.-D.}\ \bibnamefont {Song}}, \bibinfo {author}
  {\bibfnamefont {N.}~\bibnamefont {Regnault}}, \bibinfo {author}
  {\bibfnamefont {Q.}~\bibnamefont {Yang}}, \bibinfo {author} {\bibfnamefont
  {Y.}~\bibnamefont {Sun}}, \bibinfo {author} {\bibfnamefont {S.}~\bibnamefont
  {Parkin}}, \bibinfo {author} {\bibfnamefont {C.}~\bibnamefont {Felser}}, \
  and\ \bibinfo {author} {\bibfnamefont {B.~A.}\ \bibnamefont {Bernevig}},\
  }\href {https://arxiv.org/abs/2111.02433} {\bibfield  {journal} {\bibinfo
  {journal} {arXiv:2111.02433}\ } (\bibinfo {year}
  {2021}{\natexlab{a}})}\BibitemShut {NoStop}%
\bibitem [{\citenamefont {Gao}\ \emph {et~al.}(2022)\citenamefont {Gao},
  \citenamefont {Qian}, \citenamefont {Jia}, \citenamefont {Guo}, \citenamefont
  {Fang}, \citenamefont {Liu}, \citenamefont {Weng},\ and\ \citenamefont
  {Wang}}]{gao2022unconventional}%
  \BibitemOpen
  \bibfield  {author} {\bibinfo {author} {\bibfnamefont {J.}~\bibnamefont
  {Gao}}, \bibinfo {author} {\bibfnamefont {Y.}~\bibnamefont {Qian}}, \bibinfo
  {author} {\bibfnamefont {H.}~\bibnamefont {Jia}}, \bibinfo {author}
  {\bibfnamefont {Z.}~\bibnamefont {Guo}}, \bibinfo {author} {\bibfnamefont
  {Z.}~\bibnamefont {Fang}}, \bibinfo {author} {\bibfnamefont {M.}~\bibnamefont
  {Liu}}, \bibinfo {author} {\bibfnamefont {H.}~\bibnamefont {Weng}}, \ and\
  \bibinfo {author} {\bibfnamefont {Z.}~\bibnamefont {Wang}},\ }\href
  {https://www.sciencedirect.com/science/article/abs/pii/S2095927321008045}
  {\bibfield  {journal} {\bibinfo  {journal} {Science Bulletin}\ }\textbf
  {\bibinfo {volume} {67}},\ \bibinfo {pages} {598} (\bibinfo {year}
  {2022})}\BibitemShut {NoStop}%
\bibitem [{\citenamefont {Guo}\ \emph {et~al.}(2022)\citenamefont {Guo},
  \citenamefont {Deng}, \citenamefont {Xie},\ and\ \citenamefont
  {Wang}}]{guo2022quadrupole}%
  \BibitemOpen
  \bibfield  {author} {\bibinfo {author} {\bibfnamefont {Z.}~\bibnamefont
  {Guo}}, \bibinfo {author} {\bibfnamefont {J.}~\bibnamefont {Deng}}, \bibinfo
  {author} {\bibfnamefont {Y.}~\bibnamefont {Xie}}, \ and\ \bibinfo {author}
  {\bibfnamefont {Z.}~\bibnamefont {Wang}},\ }\href
  {https://www.nature.com/articles/s41535-022-00498-8} {\bibfield  {journal}
  {\bibinfo  {journal} {npj Quantum Mater.}\ }\textbf {\bibinfo {volume} {7}},\
  \bibinfo {pages} {1} (\bibinfo {year} {2022})}\BibitemShut {NoStop}%
\bibitem [{\citenamefont {Serra-Garcia}\ \emph {et~al.}(2018)\citenamefont
  {Serra-Garcia}, \citenamefont {Peri}, \citenamefont {S{\"u}sstrunk},
  \citenamefont {Bilal}, \citenamefont {Larsen}, \citenamefont {Villanueva},\
  and\ \citenamefont {Huber}}]{serra2018observation}%
  \BibitemOpen
  \bibfield  {author} {\bibinfo {author} {\bibfnamefont {M.}~\bibnamefont
  {Serra-Garcia}}, \bibinfo {author} {\bibfnamefont {V.}~\bibnamefont {Peri}},
  \bibinfo {author} {\bibfnamefont {R.}~\bibnamefont {S{\"u}sstrunk}}, \bibinfo
  {author} {\bibfnamefont {O.~R.}\ \bibnamefont {Bilal}}, \bibinfo {author}
  {\bibfnamefont {T.}~\bibnamefont {Larsen}}, \bibinfo {author} {\bibfnamefont
  {L.~G.}\ \bibnamefont {Villanueva}}, \ and\ \bibinfo {author} {\bibfnamefont
  {S.~D.}\ \bibnamefont {Huber}},\ }\href
  {https://www.nature.com/articles/nature25156} {\bibfield  {journal} {\bibinfo
   {journal} {Nature}\ }\textbf {\bibinfo {volume} {555}},\ \bibinfo {pages}
  {342} (\bibinfo {year} {2018})}\BibitemShut {NoStop}%
\bibitem [{\citenamefont {Mittal}\ \emph {et~al.}(2019)\citenamefont {Mittal},
  \citenamefont {Orre}, \citenamefont {Zhu}, \citenamefont {Gorlach},
  \citenamefont {Poddubny},\ and\ \citenamefont {Hafezi}}]{mittal2019photonic}%
  \BibitemOpen
  \bibfield  {author} {\bibinfo {author} {\bibfnamefont {S.}~\bibnamefont
  {Mittal}}, \bibinfo {author} {\bibfnamefont {V.~V.}\ \bibnamefont {Orre}},
  \bibinfo {author} {\bibfnamefont {G.}~\bibnamefont {Zhu}}, \bibinfo {author}
  {\bibfnamefont {M.~A.}\ \bibnamefont {Gorlach}}, \bibinfo {author}
  {\bibfnamefont {A.}~\bibnamefont {Poddubny}}, \ and\ \bibinfo {author}
  {\bibfnamefont {M.}~\bibnamefont {Hafezi}},\ }\href
  {https://www.nature.com/articles/s41566-019-0452-0} {\bibfield  {journal}
  {\bibinfo  {journal} {Nature Photon.}\ }\textbf {\bibinfo {volume} {13}},\
  \bibinfo {pages} {692} (\bibinfo {year} {2019})}\BibitemShut {NoStop}%
\bibitem [{\citenamefont {El~Hassan}\ \emph {et~al.}(2019)\citenamefont
  {El~Hassan}, \citenamefont {Kunst}, \citenamefont {Moritz}, \citenamefont
  {Andler}, \citenamefont {Bergholtz},\ and\ \citenamefont
  {Bourennane}}]{el2019corner}%
  \BibitemOpen
  \bibfield  {author} {\bibinfo {author} {\bibfnamefont {A.}~\bibnamefont
  {El~Hassan}}, \bibinfo {author} {\bibfnamefont {F.~K.}\ \bibnamefont
  {Kunst}}, \bibinfo {author} {\bibfnamefont {A.}~\bibnamefont {Moritz}},
  \bibinfo {author} {\bibfnamefont {G.}~\bibnamefont {Andler}}, \bibinfo
  {author} {\bibfnamefont {E.~J.}\ \bibnamefont {Bergholtz}}, \ and\ \bibinfo
  {author} {\bibfnamefont {M.}~\bibnamefont {Bourennane}},\ }\href
  {https://www.nature.com/articles/s41566-019-0519-y} {\bibfield  {journal}
  {\bibinfo  {journal} {Nature Photon.}\ }\textbf {\bibinfo {volume} {13}},\
  \bibinfo {pages} {697} (\bibinfo {year} {2019})}\BibitemShut {NoStop}%
\bibitem [{\citenamefont {Xie}\ \emph {et~al.}(2019)\citenamefont {Xie},
  \citenamefont {Su}, \citenamefont {Wang}, \citenamefont {Su}, \citenamefont
  {Shen}, \citenamefont {Zhan}, \citenamefont {Lu}, \citenamefont {Wang},\ and\
  \citenamefont {Chen}}]{PhysRevLett.122.233903}%
  \BibitemOpen
  \bibfield  {author} {\bibinfo {author} {\bibfnamefont {B.-Y.}\ \bibnamefont
  {Xie}}, \bibinfo {author} {\bibfnamefont {G.-X.}\ \bibnamefont {Su}},
  \bibinfo {author} {\bibfnamefont {H.-F.}\ \bibnamefont {Wang}}, \bibinfo
  {author} {\bibfnamefont {H.}~\bibnamefont {Su}}, \bibinfo {author}
  {\bibfnamefont {X.-P.}\ \bibnamefont {Shen}}, \bibinfo {author}
  {\bibfnamefont {P.}~\bibnamefont {Zhan}}, \bibinfo {author} {\bibfnamefont
  {M.-H.}\ \bibnamefont {Lu}}, \bibinfo {author} {\bibfnamefont {Z.-L.}\
  \bibnamefont {Wang}}, \ and\ \bibinfo {author} {\bibfnamefont {Y.-F.}\
  \bibnamefont {Chen}},\ }\href {\doibase 10.1103/PhysRevLett.122.233903}
  {\bibfield  {journal} {\bibinfo  {journal} {Phys. Rev. Lett.}\ }\textbf
  {\bibinfo {volume} {122}},\ \bibinfo {pages} {233903} (\bibinfo {year}
  {2019})}\BibitemShut {NoStop}%
\bibitem [{\citenamefont {Ma}\ \emph {et~al.}(2019)\citenamefont {Ma},
  \citenamefont {Xiao},\ and\ \citenamefont {Chan}}]{ma2019topological}%
  \BibitemOpen
  \bibfield  {author} {\bibinfo {author} {\bibfnamefont {G.}~\bibnamefont
  {Ma}}, \bibinfo {author} {\bibfnamefont {M.}~\bibnamefont {Xiao}}, \ and\
  \bibinfo {author} {\bibfnamefont {C.~T.}\ \bibnamefont {Chan}},\ }\href
  {https://www.nature.com/articles/s42254-019-0030-x} {\bibfield  {journal}
  {\bibinfo  {journal} {Nat. Rev. Phys.}\ }\textbf {\bibinfo {volume} {1}},\
  \bibinfo {pages} {281} (\bibinfo {year} {2019})}\BibitemShut {NoStop}%
\bibitem [{\citenamefont {Xue}\ \emph {et~al.}(2019)\citenamefont {Xue},
  \citenamefont {Yang}, \citenamefont {Gao}, \citenamefont {Chong},\ and\
  \citenamefont {Zhang}}]{xue2019acoustic}%
  \BibitemOpen
  \bibfield  {author} {\bibinfo {author} {\bibfnamefont {H.}~\bibnamefont
  {Xue}}, \bibinfo {author} {\bibfnamefont {Y.}~\bibnamefont {Yang}}, \bibinfo
  {author} {\bibfnamefont {F.}~\bibnamefont {Gao}}, \bibinfo {author}
  {\bibfnamefont {Y.}~\bibnamefont {Chong}}, \ and\ \bibinfo {author}
  {\bibfnamefont {B.}~\bibnamefont {Zhang}},\ }\href
  {https://www.nature.com/articles/s41563-018-0251-x} {\bibfield  {journal}
  {\bibinfo  {journal} {Nature Mater.}\ }\textbf {\bibinfo {volume} {18}},\
  \bibinfo {pages} {108} (\bibinfo {year} {2019})}\BibitemShut {NoStop}%
\bibitem [{\citenamefont {Ni}\ \emph {et~al.}(2019)\citenamefont {Ni},
  \citenamefont {Weiner}, \citenamefont {Alu},\ and\ \citenamefont
  {Khanikaev}}]{ni2019observation}%
  \BibitemOpen
  \bibfield  {author} {\bibinfo {author} {\bibfnamefont {X.}~\bibnamefont
  {Ni}}, \bibinfo {author} {\bibfnamefont {M.}~\bibnamefont {Weiner}}, \bibinfo
  {author} {\bibfnamefont {A.}~\bibnamefont {Alu}}, \ and\ \bibinfo {author}
  {\bibfnamefont {A.~B.}\ \bibnamefont {Khanikaev}},\ }\href
  {https://www.nature.com/articles/s41563-018-0252-9} {\bibfield  {journal}
  {\bibinfo  {journal} {Nature Mater.}\ }\textbf {\bibinfo {volume} {18}},\
  \bibinfo {pages} {113} (\bibinfo {year} {2019})}\BibitemShut {NoStop}%
\bibitem [{\citenamefont {Qi}\ \emph {et~al.}(2020)\citenamefont {Qi},
  \citenamefont {Qiu}, \citenamefont {Xiao}, \citenamefont {He}, \citenamefont
  {Ke},\ and\ \citenamefont {Liu}}]{PhysRevLett.124.206601}%
  \BibitemOpen
  \bibfield  {author} {\bibinfo {author} {\bibfnamefont {Y.}~\bibnamefont
  {Qi}}, \bibinfo {author} {\bibfnamefont {C.}~\bibnamefont {Qiu}}, \bibinfo
  {author} {\bibfnamefont {M.}~\bibnamefont {Xiao}}, \bibinfo {author}
  {\bibfnamefont {H.}~\bibnamefont {He}}, \bibinfo {author} {\bibfnamefont
  {M.}~\bibnamefont {Ke}}, \ and\ \bibinfo {author} {\bibfnamefont
  {Z.}~\bibnamefont {Liu}},\ }\href {\doibase 10.1103/PhysRevLett.124.206601}
  {\bibfield  {journal} {\bibinfo  {journal} {Phys. Rev. Lett.}\ }\textbf
  {\bibinfo {volume} {124}},\ \bibinfo {pages} {206601} (\bibinfo {year}
  {2020})}\BibitemShut {NoStop}%
\bibitem [{\citenamefont {Imhof}\ \emph {et~al.}(2018)\citenamefont {Imhof},
  \citenamefont {Berger}, \citenamefont {Bayer}, \citenamefont {Brehm},
  \citenamefont {Molenkamp}, \citenamefont {Kiessling}, \citenamefont
  {Schindler}, \citenamefont {Lee}, \citenamefont {Greiter}, \citenamefont
  {Neupert} \emph {et~al.}}]{imhof2018topolectrical}%
  \BibitemOpen
  \bibfield  {author} {\bibinfo {author} {\bibfnamefont {S.}~\bibnamefont
  {Imhof}}, \bibinfo {author} {\bibfnamefont {C.}~\bibnamefont {Berger}},
  \bibinfo {author} {\bibfnamefont {F.}~\bibnamefont {Bayer}}, \bibinfo
  {author} {\bibfnamefont {J.}~\bibnamefont {Brehm}}, \bibinfo {author}
  {\bibfnamefont {L.~W.}\ \bibnamefont {Molenkamp}}, \bibinfo {author}
  {\bibfnamefont {T.}~\bibnamefont {Kiessling}}, \bibinfo {author}
  {\bibfnamefont {F.}~\bibnamefont {Schindler}}, \bibinfo {author}
  {\bibfnamefont {C.~H.}\ \bibnamefont {Lee}}, \bibinfo {author} {\bibfnamefont
  {M.}~\bibnamefont {Greiter}}, \bibinfo {author} {\bibfnamefont
  {T.}~\bibnamefont {Neupert}},  \emph {et~al.},\ }\href
  {https://www.nature.com/articles/s41567-018-0246-1} {\bibfield  {journal}
  {\bibinfo  {journal} {Nature Phys.}\ }\textbf {\bibinfo {volume} {14}},\
  \bibinfo {pages} {925} (\bibinfo {year} {2018})}\BibitemShut {NoStop}%
\bibitem [{\citenamefont {Ezawa}(2018{\natexlab{c}})}]{PhysRevB.98.201402}%
  \BibitemOpen
  \bibfield  {author} {\bibinfo {author} {\bibfnamefont {M.}~\bibnamefont
  {Ezawa}},\ }\href {\doibase 10.1103/PhysRevB.98.201402} {\bibfield  {journal}
  {\bibinfo  {journal} {Phys. Rev. B}\ }\textbf {\bibinfo {volume} {98}},\
  \bibinfo {pages} {201402} (\bibinfo {year} {2018}{\natexlab{c}})}\BibitemShut
  {NoStop}%
\bibitem [{\citenamefont {Serra-Garcia}\ \emph {et~al.}(2019)\citenamefont
  {Serra-Garcia}, \citenamefont {S\"usstrunk},\ and\ \citenamefont
  {Huber}}]{PhysRevB.99.020304}%
  \BibitemOpen
  \bibfield  {author} {\bibinfo {author} {\bibfnamefont {M.}~\bibnamefont
  {Serra-Garcia}}, \bibinfo {author} {\bibfnamefont {R.}~\bibnamefont
  {S\"usstrunk}}, \ and\ \bibinfo {author} {\bibfnamefont {S.~D.}\ \bibnamefont
  {Huber}},\ }\href {\doibase 10.1103/PhysRevB.99.020304} {\bibfield  {journal}
  {\bibinfo  {journal} {Phys. Rev. B}\ }\textbf {\bibinfo {volume} {99}},\
  \bibinfo {pages} {020304} (\bibinfo {year} {2019})}\BibitemShut {NoStop}%
\bibitem [{\citenamefont {Zhang}\ \emph {et~al.}(2010)\citenamefont {Zhang},
  \citenamefont {Ren}, \citenamefont {Wang},\ and\ \citenamefont
  {Li}}]{PhysRevLett.105.225901}%
  \BibitemOpen
  \bibfield  {author} {\bibinfo {author} {\bibfnamefont {L.}~\bibnamefont
  {Zhang}}, \bibinfo {author} {\bibfnamefont {J.}~\bibnamefont {Ren}}, \bibinfo
  {author} {\bibfnamefont {J.-S.}\ \bibnamefont {Wang}}, \ and\ \bibinfo
  {author} {\bibfnamefont {B.}~\bibnamefont {Li}},\ }\href {\doibase
  10.1103/PhysRevLett.105.225901} {\bibfield  {journal} {\bibinfo  {journal}
  {Phys. Rev. Lett.}\ }\textbf {\bibinfo {volume} {105}},\ \bibinfo {pages}
  {225901} (\bibinfo {year} {2010})}\BibitemShut {NoStop}%
\bibitem [{\citenamefont {Liu}\ \emph {et~al.}(2017{\natexlab{a}})\citenamefont
  {Liu}, \citenamefont {Lian}, \citenamefont {Li}, \citenamefont {Xu},\ and\
  \citenamefont {Duan}}]{PhysRevLett.119.255901}%
  \BibitemOpen
  \bibfield  {author} {\bibinfo {author} {\bibfnamefont {Y.}~\bibnamefont
  {Liu}}, \bibinfo {author} {\bibfnamefont {C.-S.}\ \bibnamefont {Lian}},
  \bibinfo {author} {\bibfnamefont {Y.}~\bibnamefont {Li}}, \bibinfo {author}
  {\bibfnamefont {Y.}~\bibnamefont {Xu}}, \ and\ \bibinfo {author}
  {\bibfnamefont {W.}~\bibnamefont {Duan}},\ }\href {\doibase
  10.1103/PhysRevLett.119.255901} {\bibfield  {journal} {\bibinfo  {journal}
  {Phys. Rev. Lett.}\ }\textbf {\bibinfo {volume} {119}},\ \bibinfo {pages}
  {255901} (\bibinfo {year} {2017}{\natexlab{a}})}\BibitemShut {NoStop}%
\bibitem [{\citenamefont {Zhang}\ \emph {et~al.}(2018)\citenamefont {Zhang},
  \citenamefont {Song}, \citenamefont {Alexandradinata}, \citenamefont {Weng},
  \citenamefont {Fang}, \citenamefont {Lu},\ and\ \citenamefont
  {Fang}}]{PhysRevLett.120.016401}%
  \BibitemOpen
  \bibfield  {author} {\bibinfo {author} {\bibfnamefont {T.}~\bibnamefont
  {Zhang}}, \bibinfo {author} {\bibfnamefont {Z.}~\bibnamefont {Song}},
  \bibinfo {author} {\bibfnamefont {A.}~\bibnamefont {Alexandradinata}},
  \bibinfo {author} {\bibfnamefont {H.}~\bibnamefont {Weng}}, \bibinfo {author}
  {\bibfnamefont {C.}~\bibnamefont {Fang}}, \bibinfo {author} {\bibfnamefont
  {L.}~\bibnamefont {Lu}}, \ and\ \bibinfo {author} {\bibfnamefont
  {Z.}~\bibnamefont {Fang}},\ }\href {\doibase 10.1103/PhysRevLett.120.016401}
  {\bibfield  {journal} {\bibinfo  {journal} {Phys. Rev. Lett.}\ }\textbf
  {\bibinfo {volume} {120}},\ \bibinfo {pages} {016401} (\bibinfo {year}
  {2018})}\BibitemShut {NoStop}%
\bibitem [{\citenamefont {Miao}\ \emph {et~al.}(2018)\citenamefont {Miao},
  \citenamefont {Zhang}, \citenamefont {Wang}, \citenamefont {Meyers},
  \citenamefont {Said}, \citenamefont {Wang}, \citenamefont {Shi},
  \citenamefont {Weng}, \citenamefont {Fang},\ and\ \citenamefont
  {Dean}}]{PhysRevLett.121.035302}%
  \BibitemOpen
  \bibfield  {author} {\bibinfo {author} {\bibfnamefont {H.}~\bibnamefont
  {Miao}}, \bibinfo {author} {\bibfnamefont {T.~T.}\ \bibnamefont {Zhang}},
  \bibinfo {author} {\bibfnamefont {L.}~\bibnamefont {Wang}}, \bibinfo {author}
  {\bibfnamefont {D.}~\bibnamefont {Meyers}}, \bibinfo {author} {\bibfnamefont
  {A.~H.}\ \bibnamefont {Said}}, \bibinfo {author} {\bibfnamefont {Y.~L.}\
  \bibnamefont {Wang}}, \bibinfo {author} {\bibfnamefont {Y.~G.}\ \bibnamefont
  {Shi}}, \bibinfo {author} {\bibfnamefont {H.~M.}\ \bibnamefont {Weng}},
  \bibinfo {author} {\bibfnamefont {Z.}~\bibnamefont {Fang}}, \ and\ \bibinfo
  {author} {\bibfnamefont {M.~P.~M.}\ \bibnamefont {Dean}},\ }\href {\doibase
  10.1103/PhysRevLett.121.035302} {\bibfield  {journal} {\bibinfo  {journal}
  {Phys. Rev. Lett.}\ }\textbf {\bibinfo {volume} {121}},\ \bibinfo {pages}
  {035302} (\bibinfo {year} {2018})}\BibitemShut {NoStop}%
\bibitem [{\citenamefont {Zhang}\ \emph {et~al.}(2019)\citenamefont {Zhang},
  \citenamefont {Miao}, \citenamefont {Wang}, \citenamefont {Lin},
  \citenamefont {Cao}, \citenamefont {Fabbris}, \citenamefont {Said},
  \citenamefont {Liu}, \citenamefont {Lei}, \citenamefont {Fang}, \citenamefont
  {Weng},\ and\ \citenamefont {Dean}}]{PhysRevLett.123.245302}%
  \BibitemOpen
  \bibfield  {author} {\bibinfo {author} {\bibfnamefont {T.~T.}\ \bibnamefont
  {Zhang}}, \bibinfo {author} {\bibfnamefont {H.}~\bibnamefont {Miao}},
  \bibinfo {author} {\bibfnamefont {Q.}~\bibnamefont {Wang}}, \bibinfo {author}
  {\bibfnamefont {J.~Q.}\ \bibnamefont {Lin}}, \bibinfo {author} {\bibfnamefont
  {Y.}~\bibnamefont {Cao}}, \bibinfo {author} {\bibfnamefont {G.}~\bibnamefont
  {Fabbris}}, \bibinfo {author} {\bibfnamefont {A.~H.}\ \bibnamefont {Said}},
  \bibinfo {author} {\bibfnamefont {X.}~\bibnamefont {Liu}}, \bibinfo {author}
  {\bibfnamefont {H.~C.}\ \bibnamefont {Lei}}, \bibinfo {author} {\bibfnamefont
  {Z.}~\bibnamefont {Fang}}, \bibinfo {author} {\bibfnamefont {H.~M.}\
  \bibnamefont {Weng}}, \ and\ \bibinfo {author} {\bibfnamefont {M.~P.~M.}\
  \bibnamefont {Dean}},\ }\href {\doibase 10.1103/PhysRevLett.123.245302}
  {\bibfield  {journal} {\bibinfo  {journal} {Phys. Rev. Lett.}\ }\textbf
  {\bibinfo {volume} {123}},\ \bibinfo {pages} {245302} (\bibinfo {year}
  {2019})}\BibitemShut {NoStop}%
\bibitem [{\citenamefont {Xia}\ \emph {et~al.}(2019)\citenamefont {Xia},
  \citenamefont {Wang}, \citenamefont {Chen}, \citenamefont {Zhao},\ and\
  \citenamefont {Xu}}]{PhysRevLett.123.065501}%
  \BibitemOpen
  \bibfield  {author} {\bibinfo {author} {\bibfnamefont {B.~W.}\ \bibnamefont
  {Xia}}, \bibinfo {author} {\bibfnamefont {R.}~\bibnamefont {Wang}}, \bibinfo
  {author} {\bibfnamefont {Z.~J.}\ \bibnamefont {Chen}}, \bibinfo {author}
  {\bibfnamefont {Y.~J.}\ \bibnamefont {Zhao}}, \ and\ \bibinfo {author}
  {\bibfnamefont {H.}~\bibnamefont {Xu}},\ }\href {\doibase
  10.1103/PhysRevLett.123.065501} {\bibfield  {journal} {\bibinfo  {journal}
  {Phys. Rev. Lett.}\ }\textbf {\bibinfo {volume} {123}},\ \bibinfo {pages}
  {065501} (\bibinfo {year} {2019})}\BibitemShut {NoStop}%
\bibitem [{\citenamefont {Long}\ \emph {et~al.}(2020)\citenamefont {Long},
  \citenamefont {Ren},\ and\ \citenamefont {Chen}}]{PhysRevLett.124.185501}%
  \BibitemOpen
  \bibfield  {author} {\bibinfo {author} {\bibfnamefont {Y.}~\bibnamefont
  {Long}}, \bibinfo {author} {\bibfnamefont {J.}~\bibnamefont {Ren}}, \ and\
  \bibinfo {author} {\bibfnamefont {H.}~\bibnamefont {Chen}},\ }\href {\doibase
  10.1103/PhysRevLett.124.185501} {\bibfield  {journal} {\bibinfo  {journal}
  {Phys. Rev. Lett.}\ }\textbf {\bibinfo {volume} {124}},\ \bibinfo {pages}
  {185501} (\bibinfo {year} {2020})}\BibitemShut {NoStop}%
\bibitem [{\citenamefont {Liu}\ \emph {et~al.}(2020)\citenamefont {Liu},
  \citenamefont {Chen},\ and\ \citenamefont {Xu}}]{liu2020topological}%
  \BibitemOpen
  \bibfield  {author} {\bibinfo {author} {\bibfnamefont {Y.}~\bibnamefont
  {Liu}}, \bibinfo {author} {\bibfnamefont {X.}~\bibnamefont {Chen}}, \ and\
  \bibinfo {author} {\bibfnamefont {Y.}~\bibnamefont {Xu}},\ }\href
  {https://onlinelibrary.wiley.com/doi/10.1002/adfm.201904784} {\bibfield
  {journal} {\bibinfo  {journal} {Adv. Funct. Mater.}\ }\textbf {\bibinfo
  {volume} {30}},\ \bibinfo {pages} {1904784} (\bibinfo {year}
  {2020})}\BibitemShut {NoStop}%
\bibitem [{\citenamefont {Wang}\ \emph {et~al.}(2020)\citenamefont {Wang},
  \citenamefont {Xia}, \citenamefont {Chen}, \citenamefont {Zheng},
  \citenamefont {Zhao},\ and\ \citenamefont {Xu}}]{PhysRevLett.124.105303}%
  \BibitemOpen
  \bibfield  {author} {\bibinfo {author} {\bibfnamefont {R.}~\bibnamefont
  {Wang}}, \bibinfo {author} {\bibfnamefont {B.~W.}\ \bibnamefont {Xia}},
  \bibinfo {author} {\bibfnamefont {Z.~J.}\ \bibnamefont {Chen}}, \bibinfo
  {author} {\bibfnamefont {B.~B.}\ \bibnamefont {Zheng}}, \bibinfo {author}
  {\bibfnamefont {Y.~J.}\ \bibnamefont {Zhao}}, \ and\ \bibinfo {author}
  {\bibfnamefont {H.}~\bibnamefont {Xu}},\ }\href {\doibase
  10.1103/PhysRevLett.124.105303} {\bibfield  {journal} {\bibinfo  {journal}
  {Phys. Rev. Lett.}\ }\textbf {\bibinfo {volume} {124}},\ \bibinfo {pages}
  {105303} (\bibinfo {year} {2020})}\BibitemShut {NoStop}%
\bibitem [{\citenamefont {Li}\ \emph {et~al.}(2020)\citenamefont {Li},
  \citenamefont {Wang}, \citenamefont {Liu}, \citenamefont {Li}, \citenamefont
  {Zhang},\ and\ \citenamefont {Chen}}]{PhysRevB.101.081403}%
  \BibitemOpen
  \bibfield  {author} {\bibinfo {author} {\bibfnamefont {J.}~\bibnamefont
  {Li}}, \bibinfo {author} {\bibfnamefont {L.}~\bibnamefont {Wang}}, \bibinfo
  {author} {\bibfnamefont {J.}~\bibnamefont {Liu}}, \bibinfo {author}
  {\bibfnamefont {R.}~\bibnamefont {Li}}, \bibinfo {author} {\bibfnamefont
  {Z.}~\bibnamefont {Zhang}}, \ and\ \bibinfo {author} {\bibfnamefont {X.-Q.}\
  \bibnamefont {Chen}},\ }\href {\doibase 10.1103/PhysRevB.101.081403}
  {\bibfield  {journal} {\bibinfo  {journal} {Phys. Rev. B}\ }\textbf {\bibinfo
  {volume} {101}},\ \bibinfo {pages} {081403} (\bibinfo {year}
  {2020})}\BibitemShut {NoStop}%
\bibitem [{\citenamefont {Li}\ \emph {et~al.}(2021)\citenamefont {Li},
  \citenamefont {Liu}, \citenamefont {Baronett}, \citenamefont {Liu},
  \citenamefont {Wang}, \citenamefont {Li}, \citenamefont {Chen}, \citenamefont
  {Li}, \citenamefont {Zhu},\ and\ \citenamefont {Chen}}]{li2021computation}%
  \BibitemOpen
  \bibfield  {author} {\bibinfo {author} {\bibfnamefont {J.}~\bibnamefont
  {Li}}, \bibinfo {author} {\bibfnamefont {J.}~\bibnamefont {Liu}}, \bibinfo
  {author} {\bibfnamefont {S.~A.}\ \bibnamefont {Baronett}}, \bibinfo {author}
  {\bibfnamefont {M.}~\bibnamefont {Liu}}, \bibinfo {author} {\bibfnamefont
  {L.}~\bibnamefont {Wang}}, \bibinfo {author} {\bibfnamefont {R.}~\bibnamefont
  {Li}}, \bibinfo {author} {\bibfnamefont {Y.}~\bibnamefont {Chen}}, \bibinfo
  {author} {\bibfnamefont {D.}~\bibnamefont {Li}}, \bibinfo {author}
  {\bibfnamefont {Q.}~\bibnamefont {Zhu}}, \ and\ \bibinfo {author}
  {\bibfnamefont {X.-Q.}\ \bibnamefont {Chen}},\ }\href
  {https://www.nature.com/articles/s41467-021-21293-2} {\bibfield  {journal}
  {\bibinfo  {journal} {Nat. Commun.}\ }\textbf {\bibinfo {volume} {12}},\
  \bibinfo {pages} {1} (\bibinfo {year} {2021})}\BibitemShut {NoStop}%
\bibitem [{\citenamefont {Zhu}\ \emph {et~al.}(2022)\citenamefont {Zhu},
  \citenamefont {Wu}, \citenamefont {Zhao}, \citenamefont {Chen}, \citenamefont
  {Wang}, \citenamefont {Sheng}, \citenamefont {Zhang}, \citenamefont {Zhao},\
  and\ \citenamefont {Yang}}]{PhysRevB.105.085123}%
  \BibitemOpen
  \bibfield  {author} {\bibinfo {author} {\bibfnamefont {J.}~\bibnamefont
  {Zhu}}, \bibinfo {author} {\bibfnamefont {W.}~\bibnamefont {Wu}}, \bibinfo
  {author} {\bibfnamefont {J.}~\bibnamefont {Zhao}}, \bibinfo {author}
  {\bibfnamefont {C.}~\bibnamefont {Chen}}, \bibinfo {author} {\bibfnamefont
  {Q.}~\bibnamefont {Wang}}, \bibinfo {author} {\bibfnamefont {X.-L.}\
  \bibnamefont {Sheng}}, \bibinfo {author} {\bibfnamefont {L.}~\bibnamefont
  {Zhang}}, \bibinfo {author} {\bibfnamefont {Y.~X.}\ \bibnamefont {Zhao}}, \
  and\ \bibinfo {author} {\bibfnamefont {S.~A.}\ \bibnamefont {Yang}},\ }\href
  {\doibase 10.1103/PhysRevB.105.085123} {\bibfield  {journal} {\bibinfo
  {journal} {Phys. Rev. B}\ }\textbf {\bibinfo {volume} {105}},\ \bibinfo
  {pages} {085123} (\bibinfo {year} {2022})}\BibitemShut {NoStop}%
\bibitem [{\citenamefont {Peri}\ \emph {et~al.}(2021)\citenamefont {Peri},
  \citenamefont {Song}, \citenamefont {Bernevig},\ and\ \citenamefont
  {Huber}}]{PhysRevLett.126.027002}%
  \BibitemOpen
  \bibfield  {author} {\bibinfo {author} {\bibfnamefont {V.}~\bibnamefont
  {Peri}}, \bibinfo {author} {\bibfnamefont {Z.-D.}\ \bibnamefont {Song}},
  \bibinfo {author} {\bibfnamefont {B.~A.}\ \bibnamefont {Bernevig}}, \ and\
  \bibinfo {author} {\bibfnamefont {S.~D.}\ \bibnamefont {Huber}},\ }\href
  {\doibase 10.1103/PhysRevLett.126.027002} {\bibfield  {journal} {\bibinfo
  {journal} {Phys. Rev. Lett.}\ }\textbf {\bibinfo {volume} {126}},\ \bibinfo
  {pages} {027002} (\bibinfo {year} {2021})}\BibitemShut {NoStop}%
\bibitem [{\citenamefont {Zhang}\ \emph {et~al.}(2016)\citenamefont {Zhang},
  \citenamefont {Xiang},\ and\ \citenamefont {Li}}]{zhang2016theoretical}%
  \BibitemOpen
  \bibfield  {author} {\bibinfo {author} {\bibfnamefont {W.-B.}\ \bibnamefont
  {Zhang}}, \bibinfo {author} {\bibfnamefont {L.-J.}\ \bibnamefont {Xiang}}, \
  and\ \bibinfo {author} {\bibfnamefont {H.-B.}\ \bibnamefont {Li}},\ }\href
  {https://pubs.rsc.org/en/content/articlelanding/2016/ta/c6ta06806e/unauth}
  {\bibfield  {journal} {\bibinfo  {journal} {J. Mater. Chem.A}\ }\textbf
  {\bibinfo {volume} {4}},\ \bibinfo {pages} {19086} (\bibinfo {year}
  {2016})}\BibitemShut {NoStop}%
\bibitem [{\citenamefont {Liu}\ \emph {et~al.}(2017{\natexlab{b}})\citenamefont
  {Liu}, \citenamefont {Lu}, \citenamefont {Wu}, \citenamefont {Luo},
  \citenamefont {Cheng}, \citenamefont {Wang}, \citenamefont {Wang},
  \citenamefont {Wang}, \citenamefont {Liu},\ and\ \citenamefont
  {Cho}}]{liu2017electronic}%
  \BibitemOpen
  \bibfield  {author} {\bibinfo {author} {\bibfnamefont {P.}~\bibnamefont
  {Liu}}, \bibinfo {author} {\bibfnamefont {F.}~\bibnamefont {Lu}}, \bibinfo
  {author} {\bibfnamefont {M.}~\bibnamefont {Wu}}, \bibinfo {author}
  {\bibfnamefont {X.}~\bibnamefont {Luo}}, \bibinfo {author} {\bibfnamefont
  {Y.}~\bibnamefont {Cheng}}, \bibinfo {author} {\bibfnamefont {X.-W.}\
  \bibnamefont {Wang}}, \bibinfo {author} {\bibfnamefont {W.}~\bibnamefont
  {Wang}}, \bibinfo {author} {\bibfnamefont {W.-H.}\ \bibnamefont {Wang}},
  \bibinfo {author} {\bibfnamefont {H.}~\bibnamefont {Liu}}, \ and\ \bibinfo
  {author} {\bibfnamefont {K.}~\bibnamefont {Cho}},\ }\href
  {https://pubs.rsc.org/en/content/articlelanding/2017/tc/c7tc03003g/unauth}
  {\bibfield  {journal} {\bibinfo  {journal} {J. Mater. Chem.C}\ }\textbf
  {\bibinfo {volume} {5}},\ \bibinfo {pages} {9066} (\bibinfo {year}
  {2017}{\natexlab{b}})}\BibitemShut {NoStop}%
\bibitem [{\citenamefont {Li}\ \emph {et~al.}(2017)\citenamefont {Li},
  \citenamefont {Guan}, \citenamefont {Wang}, \citenamefont {Cheng},
  \citenamefont {Ai}, \citenamefont {Yao}, \citenamefont {Chen}, \citenamefont
  {Zhang}, \citenamefont {Duan},\ and\ \citenamefont {Duan}}]{li2017synthesis}%
  \BibitemOpen
  \bibfield  {author} {\bibinfo {author} {\bibfnamefont {J.}~\bibnamefont
  {Li}}, \bibinfo {author} {\bibfnamefont {X.}~\bibnamefont {Guan}}, \bibinfo
  {author} {\bibfnamefont {C.}~\bibnamefont {Wang}}, \bibinfo {author}
  {\bibfnamefont {H.-C.}\ \bibnamefont {Cheng}}, \bibinfo {author}
  {\bibfnamefont {R.}~\bibnamefont {Ai}}, \bibinfo {author} {\bibfnamefont
  {K.}~\bibnamefont {Yao}}, \bibinfo {author} {\bibfnamefont {P.}~\bibnamefont
  {Chen}}, \bibinfo {author} {\bibfnamefont {Z.}~\bibnamefont {Zhang}},
  \bibinfo {author} {\bibfnamefont {X.}~\bibnamefont {Duan}}, \ and\ \bibinfo
  {author} {\bibfnamefont {X.}~\bibnamefont {Duan}},\ }\href
  {https://onlinelibrary.wiley.com/doi/abs/10.1002/smll.201701034} {\bibfield
  {journal} {\bibinfo  {journal} {Small}\ }\textbf {\bibinfo {volume} {13}},\
  \bibinfo {pages} {1701034} (\bibinfo {year} {2017})}\BibitemShut {NoStop}%
\bibitem [{SUP()}]{SUPP}%
  \BibitemOpen
  \href@noop {} {\bibinfo  {journal} {See the Supplemental Material for the
  details of calculations, lattice parameters for different MX$_3$ monolayers,
  the band representation of phonoon spectrum of BiI$_3$, and the effect of
  atomic terminal of nano-disk on the corner modes, which includes Refs.
  ~\cite{PhysRev.140.A1133,PhysRevB.54.11169,PhysRevB.59.1758,PhysRevLett.45.566,PhysRevLett.77.3865,PhysRevLett.78.1396,PhysRevB.13.5188,togo2015first,zhang2022phonon,wu2018wanniertools,mccollum1990layered,de1992structure,templeton1954crystal,fjellvaag1994crystal,zhao2020equivariant}}\
  }\BibitemShut {NoStop}%
\bibitem [{\citenamefont {Kresse}\ and\ \citenamefont
  {Furthm\"uller}(1996)}]{PhysRevB.54.11169}%
  \BibitemOpen
\bibfield  {journal} {  }\bibfield  {author} {\bibinfo {author} {\bibfnamefont
  {G.}~\bibnamefont {Kresse}}\ and\ \bibinfo {author} {\bibfnamefont
  {J.}~\bibnamefont {Furthm\"uller}},\ }\href {\doibase
  10.1103/PhysRevB.54.11169} {\bibfield  {journal} {\bibinfo  {journal} {Phys.
  Rev. B}\ }\textbf {\bibinfo {volume} {54}},\ \bibinfo {pages} {11169}
  (\bibinfo {year} {1996})}\BibitemShut {NoStop}%
\bibitem [{\citenamefont {Togo}\ and\ \citenamefont
  {Tanaka}(2015)}]{togo2015first}%
  \BibitemOpen
  \bibfield  {author} {\bibinfo {author} {\bibfnamefont {A.}~\bibnamefont
  {Togo}}\ and\ \bibinfo {author} {\bibfnamefont {I.}~\bibnamefont {Tanaka}},\
  }\href {https://www.sciencedirect.com/science/article/pii/S1359646215003127}
  {\bibfield  {journal} {\bibinfo  {journal} {Scr. Mater.}\ }\textbf {\bibinfo
  {volume} {108}},\ \bibinfo {pages} {1} (\bibinfo {year} {2015})}\BibitemShut
  {NoStop}%
\bibitem [{\citenamefont {Wang}\ \emph {et~al.}(2016)\citenamefont {Wang},
  \citenamefont {Alexandradinata}, \citenamefont {Cava},\ and\ \citenamefont
  {Bernevig}}]{wang2016hourglass}%
  \BibitemOpen
  \bibfield  {author} {\bibinfo {author} {\bibfnamefont {Z.}~\bibnamefont
  {Wang}}, \bibinfo {author} {\bibfnamefont {A.}~\bibnamefont
  {Alexandradinata}}, \bibinfo {author} {\bibfnamefont {R.~J.}\ \bibnamefont
  {Cava}}, \ and\ \bibinfo {author} {\bibfnamefont {B.~A.}\ \bibnamefont
  {Bernevig}},\ }\href {https://www.nature.com/articles/nature17410} {\bibfield
   {journal} {\bibinfo  {journal} {Nature}\ }\textbf {\bibinfo {volume}
  {532}},\ \bibinfo {pages} {189} (\bibinfo {year} {2016})}\BibitemShut
  {NoStop}%
\bibitem [{\citenamefont {Xu}\ \emph {et~al.}(2021{\natexlab{b}})\citenamefont
  {Xu}, \citenamefont {Elcoro}, \citenamefont {Song}, \citenamefont
  {Vergniory}, \citenamefont {Felser}, \citenamefont {Parkin}, \citenamefont
  {Regnault}, \citenamefont {Ma{\~n}es},\ and\ \citenamefont
  {Bernevig}}]{xu2021filling}%
  \BibitemOpen
  \bibfield  {author} {\bibinfo {author} {\bibfnamefont {Y.}~\bibnamefont
  {Xu}}, \bibinfo {author} {\bibfnamefont {L.}~\bibnamefont {Elcoro}}, \bibinfo
  {author} {\bibfnamefont {Z.-D.}\ \bibnamefont {Song}}, \bibinfo {author}
  {\bibfnamefont {M.}~\bibnamefont {Vergniory}}, \bibinfo {author}
  {\bibfnamefont {C.}~\bibnamefont {Felser}}, \bibinfo {author} {\bibfnamefont
  {S.~S.}\ \bibnamefont {Parkin}}, \bibinfo {author} {\bibfnamefont
  {N.}~\bibnamefont {Regnault}}, \bibinfo {author} {\bibfnamefont {J.~L.}\
  \bibnamefont {Ma{\~n}es}}, \ and\ \bibinfo {author} {\bibfnamefont {B.~A.}\
  \bibnamefont {Bernevig}},\ }\href {https://arxiv.org/abs/2106.10276}
  {\bibfield  {journal} {\bibinfo  {journal} {arXiv:2106.10276}\ } (\bibinfo
  {year} {2021}{\natexlab{b}})}\BibitemShut {NoStop}%
\bibitem [{\citenamefont {Zhao}(2020)}]{zhao2020equivariant}%
  \BibitemOpen
  \bibfield  {author} {\bibinfo {author} {\bibfnamefont {Y.}~\bibnamefont
  {Zhao}},\ }\href
  {https://link.springer.com/article/10.1007/s11467-019-0943-y} {\bibfield
  {journal} {\bibinfo  {journal} {Front. Phys.}\ }\textbf {\bibinfo {volume}
  {15}},\ \bibinfo {pages} {1} (\bibinfo {year} {2020})}\BibitemShut {NoStop}%
\bibitem [{\citenamefont {Kohn}\ and\ \citenamefont
  {Sham}(1965)}]{PhysRev.140.A1133}%
  \BibitemOpen
  \bibfield  {author} {\bibinfo {author} {\bibfnamefont {W.}~\bibnamefont
  {Kohn}}\ and\ \bibinfo {author} {\bibfnamefont {L.~J.}\ \bibnamefont
  {Sham}},\ }\href {\doibase 10.1103/PhysRev.140.A1133} {\bibfield  {journal}
  {\bibinfo  {journal} {Phys. Rev.}\ }\textbf {\bibinfo {volume} {140}},\
  \bibinfo {pages} {A1133} (\bibinfo {year} {1965})}\BibitemShut {NoStop}%
\bibitem [{\citenamefont {Kresse}\ and\ \citenamefont
  {Joubert}(1999)}]{PhysRevB.59.1758}%
  \BibitemOpen
  \bibfield  {author} {\bibinfo {author} {\bibfnamefont {G.}~\bibnamefont
  {Kresse}}\ and\ \bibinfo {author} {\bibfnamefont {D.}~\bibnamefont
  {Joubert}},\ }\href {\doibase 10.1103/PhysRevB.59.1758} {\bibfield  {journal}
  {\bibinfo  {journal} {Phys. Rev. B}\ }\textbf {\bibinfo {volume} {59}},\
  \bibinfo {pages} {1758} (\bibinfo {year} {1999})}\BibitemShut {NoStop}%
\bibitem [{\citenamefont {Ceperley}\ and\ \citenamefont
  {Alder}(1980)}]{PhysRevLett.45.566}%
  \BibitemOpen
  \bibfield  {author} {\bibinfo {author} {\bibfnamefont {D.~M.}\ \bibnamefont
  {Ceperley}}\ and\ \bibinfo {author} {\bibfnamefont {B.~J.}\ \bibnamefont
  {Alder}},\ }\href {\doibase 10.1103/PhysRevLett.45.566} {\bibfield  {journal}
  {\bibinfo  {journal} {Phys. Rev. Lett.}\ }\textbf {\bibinfo {volume} {45}},\
  \bibinfo {pages} {566} (\bibinfo {year} {1980})}\BibitemShut {NoStop}%
\bibitem [{\citenamefont {Perdew}\ \emph {et~al.}(1996)\citenamefont {Perdew},
  \citenamefont {Burke},\ and\ \citenamefont
  {Ernzerhof}}]{PhysRevLett.77.3865}%
  \BibitemOpen
  \bibfield  {author} {\bibinfo {author} {\bibfnamefont {J.~P.}\ \bibnamefont
  {Perdew}}, \bibinfo {author} {\bibfnamefont {K.}~\bibnamefont {Burke}}, \
  and\ \bibinfo {author} {\bibfnamefont {M.}~\bibnamefont {Ernzerhof}},\ }\href
  {\doibase 10.1103/PhysRevLett.77.3865} {\bibfield  {journal} {\bibinfo
  {journal} {Phys. Rev. Lett.}\ }\textbf {\bibinfo {volume} {77}},\ \bibinfo
  {pages} {3865} (\bibinfo {year} {1996})}\BibitemShut {NoStop}%
\bibitem [{\citenamefont {Perdew}\ \emph {et~al.}(1997)\citenamefont {Perdew},
  \citenamefont {Burke},\ and\ \citenamefont
  {Ernzerhof}}]{PhysRevLett.78.1396}%
  \BibitemOpen
  \bibfield  {author} {\bibinfo {author} {\bibfnamefont {J.~P.}\ \bibnamefont
  {Perdew}}, \bibinfo {author} {\bibfnamefont {K.}~\bibnamefont {Burke}}, \
  and\ \bibinfo {author} {\bibfnamefont {M.}~\bibnamefont {Ernzerhof}},\ }\href
  {\doibase 10.1103/PhysRevLett.78.1396} {\bibfield  {journal} {\bibinfo
  {journal} {Phys. Rev. Lett.}\ }\textbf {\bibinfo {volume} {78}},\ \bibinfo
  {pages} {1396} (\bibinfo {year} {1997})}\BibitemShut {NoStop}%
\bibitem [{\citenamefont {Monkhorst}\ and\ \citenamefont
  {Pack}(1976)}]{PhysRevB.13.5188}%
  \BibitemOpen
  \bibfield  {author} {\bibinfo {author} {\bibfnamefont {H.~J.}\ \bibnamefont
  {Monkhorst}}\ and\ \bibinfo {author} {\bibfnamefont {J.~D.}\ \bibnamefont
  {Pack}},\ }\href {\doibase 10.1103/PhysRevB.13.5188} {\bibfield  {journal}
  {\bibinfo  {journal} {Phys. Rev. B}\ }\textbf {\bibinfo {volume} {13}},\
  \bibinfo {pages} {5188} (\bibinfo {year} {1976})}\BibitemShut {NoStop}%
\bibitem [{\citenamefont {Zhang}\ \emph {et~al.}(2022)\citenamefont {Zhang},
  \citenamefont {Yu}, \citenamefont {Liu},\ and\ \citenamefont
  {Yao}}]{zhang2022phonon}%
  \BibitemOpen
  \bibfield  {author} {\bibinfo {author} {\bibfnamefont {Z.}~\bibnamefont
  {Zhang}}, \bibinfo {author} {\bibfnamefont {Z.-M.}\ \bibnamefont {Yu}},
  \bibinfo {author} {\bibfnamefont {G.-B.}\ \bibnamefont {Liu}}, \ and\
  \bibinfo {author} {\bibfnamefont {Y.}~\bibnamefont {Yao}},\ }\href
  {https://arxiv.org/abs/2201.11350} {\bibfield  {journal} {\bibinfo  {journal}
  {arXiv:2201.11350}\ } (\bibinfo {year} {2022})}\BibitemShut {NoStop}%
\bibitem [{\citenamefont {Wu}\ \emph {et~al.}(2018)\citenamefont {Wu},
  \citenamefont {Zhang}, \citenamefont {Song}, \citenamefont {Troyer},\ and\
  \citenamefont {Soluyanov}}]{wu2018wanniertools}%
  \BibitemOpen
  \bibfield  {author} {\bibinfo {author} {\bibfnamefont {Q.}~\bibnamefont
  {Wu}}, \bibinfo {author} {\bibfnamefont {S.}~\bibnamefont {Zhang}}, \bibinfo
  {author} {\bibfnamefont {H.-F.}\ \bibnamefont {Song}}, \bibinfo {author}
  {\bibfnamefont {M.}~\bibnamefont {Troyer}}, \ and\ \bibinfo {author}
  {\bibfnamefont {A.~A.}\ \bibnamefont {Soluyanov}},\ }\href
  {https://www.sciencedirect.com/science/article/abs/pii/S0010465517303442}
  {\bibfield  {journal} {\bibinfo  {journal} {Comput. Phys. Commun.}\ }\textbf
  {\bibinfo {volume} {224}},\ \bibinfo {pages} {405} (\bibinfo {year}
  {2018})}\BibitemShut {NoStop}%
\bibitem [{\citenamefont {McCollum}\ \emph {et~al.}(1990)\citenamefont
  {McCollum}, \citenamefont {Dudis}, \citenamefont {Lachgar},\ and\
  \citenamefont {Corbett}}]{mccollum1990layered}%
  \BibitemOpen
  \bibfield  {author} {\bibinfo {author} {\bibfnamefont {B.}~\bibnamefont
  {McCollum}}, \bibinfo {author} {\bibfnamefont {D.}~\bibnamefont {Dudis}},
  \bibinfo {author} {\bibfnamefont {A.}~\bibnamefont {Lachgar}}, \ and\
  \bibinfo {author} {\bibfnamefont {J.~D.}\ \bibnamefont {Corbett}},\ }\href
  {https://pubs.acs.org/doi/abs/10.1021/ic00335a054} {\bibfield  {journal}
  {\bibinfo  {journal} {Inorg. Chem.}\ }\textbf {\bibinfo {volume} {29}},\
  \bibinfo {pages} {2030} (\bibinfo {year} {1990})}\BibitemShut {NoStop}%
\bibitem [{\citenamefont {de~Barros~Marques}\ \emph {et~al.}(1992)\citenamefont
  {de~Barros~Marques}, \citenamefont {Marques},\ and\ \citenamefont
  {Rodrigues}}]{de1992structure}%
  \BibitemOpen
  \bibfield  {author} {\bibinfo {author} {\bibfnamefont {M.}~\bibnamefont
  {de~Barros~Marques}}, \bibinfo {author} {\bibfnamefont {M.~A.}\ \bibnamefont
  {Marques}}, \ and\ \bibinfo {author} {\bibfnamefont {J.~R.}\ \bibnamefont
  {Rodrigues}},\ }\href
  {https://iopscience.iop.org/article/10.1088/0953-8984/4/38/004/meta}
  {\bibfield  {journal} {\bibinfo  {journal} {J. Phys. Condens. Matter}\
  }\textbf {\bibinfo {volume} {4}},\ \bibinfo {pages} {7679} (\bibinfo {year}
  {1992})}\BibitemShut {NoStop}%
\bibitem [{\citenamefont {Templeton}\ and\ \citenamefont
  {Carter}(1954)}]{templeton1954crystal}%
  \BibitemOpen
  \bibfield  {author} {\bibinfo {author} {\bibfnamefont {D.}~\bibnamefont
  {Templeton}}\ and\ \bibinfo {author} {\bibfnamefont {G.~F.}\ \bibnamefont
  {Carter}},\ }\href {https://pubs.acs.org/doi/pdf/10.1021/j150521a002}
  {\bibfield  {journal} {\bibinfo  {journal} {J. Phys. Chem.}\ }\textbf
  {\bibinfo {volume} {58}},\ \bibinfo {pages} {940} (\bibinfo {year}
  {1954})}\BibitemShut {NoStop}%
\bibitem [{\citenamefont {Fjellv{\aa}g}\ and\ \citenamefont
  {Karen}(1994)}]{fjellvaag1994crystal}%
  \BibitemOpen
  \bibfield  {author} {\bibinfo {author} {\bibfnamefont {H.}~\bibnamefont
  {Fjellv{\aa}g}}\ and\ \bibinfo {author} {\bibfnamefont {P.}~\bibnamefont
  {Karen}},\ }\href
  {https://www.duo.uio.no/bitstream/handle/10852/59528/1/1994%2BScCl3.pdf}
  {\bibfield  {journal} {\bibinfo  {journal} {Acta Chem. Scand.}\ }\textbf
  {\bibinfo {volume} {48}},\ \bibinfo {pages} {294} (\bibinfo {year}
  {1994})}\BibitemShut {NoStop}%
\end{thebibliography}%

\pagebreak
\widetext
\clearpage

	\setcounter{equation}{0}
	\setcounter{figure}{0}
	\setcounter{table}{0}
	\makeatletter
	\renewcommand{\figurename}{FIG.}
	\renewcommand{\theequation}{S\arabic{equation}}
	\renewcommand{\thetable}{S\arabic{table}}
	\renewcommand{\thefigure}{S\arabic{figure}}

	\begin{center}
		\textbf{Supplemental materials of `` Phononic Obstructed Atomic Insulators with Robust Corner Modes
"}
	\end{center}

 \section{I. Computational Methods}\label{cal}
Density functional theory (DFT)~\cite{PhysRev.140.A1133} calculations are performed utilizing the Vienna $ab$ $Initio$ simulation package (VASP)~\cite{PhysRevB.54.11169} with the projector-augmented wave potential method~\cite{PhysRevB.59.1758,PhysRevLett.45.566} and the generalized gradient approximation with the Perdew-Burke-Ernzerhof as the exchange-correlation functional is applied~\cite{PhysRevLett.77.3865,PhysRevLett.78.1396}. 
During the structural relaxation, kinetic energy cutoff of plane-wave basis set is 300 eV for all $MX_3$ systems expect for ScCl$_3$ (320~eV) and YCl$_3$ (350~eV), while 1.2$\times$cutoff are used during phonon calculations, and the Brillouin zone (BZ) are sampled by 8$\times$8$\times$1 Gamma-centered Monkhorst-Pack grid~\cite{PhysRevB.13.5188}. 
Structure parameters of the ground states are obtained using the energy convergence criteria of $10^{-8}$~eV. 
Supercell size  for the calculation of force constants are 3$\times$3$\times$1 (or 4$\times$4$\times$1) with careful convergence tests, and the corresponding k-mesh applied is 3$\times$3$\times$1 (or 2$\times$2$\times$1). 
We calculate the force constants with finite displacement method using VASP. 
The phonon dispersion is obtained using Phonopy package~\cite{togo2015first}. 
The band representations of the phonon branches are obtained by PhononIrep package~\cite{zhang2022phonon}. 
To compute the phonon spectra for different nano-disk geometries, we use WannierTools package ~\cite{wu2018wanniertools} to build a phononic tight-binding (TB) model for periodic $MX_3$ monolayers. We built models for different nano-disks based on the TB model of $MX_3$ monolayers. 

\begin{table}[b]
\caption{
The structure parameters of $MX_3$ monolayers. 
The first column is the name of the compunds. 
The second column gives the lattice constant $a$. 
The length of the shortest $M$-$X$ bond, \textit{i.e.}, $d(M$-$X)$ is listed in the third column. 
The angle between the $M$-$X$ bond and the direction normal to the plane, \textit{i.e.}, $\theta$ is given in the fourth column. 
The experimental in-plane lattice constant of the bulk $MX_3$ is shown in the last 
column~\cite{mccollum1990layered,de1992structure,templeton1954crystal,fjellvaag1994crystal}.}
\vspace{0.2cm}
\renewcommand\arraystretch{1.8}
\setlength{\tabcolsep}{6mm}{
\begin{tabular}{ccccc}
\hline 
\hline
 & $a$ & $d(M$-$X)$ & $\theta$ & Experiments\tabularnewline
\hline 
BiI$_3$    & 7.83~Å  & 3.12~Å & 54.26º & 7.52~Å \tabularnewline
SbI$_3$    & 7.67~Å  & 3.06~Å & 54.22º & 7.50~Å \tabularnewline
ScI$_3$    & 7.32~Å  & 2.89~Å & 54.59º & 7.15~Å \tabularnewline
YI$_3$     & 7.71~Å  & 3.05~Å & 54.51º & 7.49~Å \tabularnewline
AsI$_3$    & 7.29~Å  & 2.89~Å & 54.45º & 7.03~Å \tabularnewline
ScBr$_3$   & 6.77~Å  & 2.64~Å & 54.81º & 6.66~Å \tabularnewline
YBr$_3$    & 7.24~Å  & 2.81~Å & 54.93º & 7.07~Å \tabularnewline
BiBr$_3$   & 7.38~Å  & 2.91~Å & 54.62º & 7.24~Å \tabularnewline
YCl$_3$    & 6.97~Å  & 2.66~Å & 55.66º & 6.90~Å \tabularnewline
ScCl$_3$   & 6.44~Å  & 2.49~Å & 55.14º & 6.38~Å \tabularnewline
\hline 
\hline
\end{tabular}}
\label{Tab-Str}
\end{table}

\section{II. The structure parameters and phonon spectra of $MX_3$}

In this section, we show the structure parameters and phonon spectra of $MX_3$ monolayers. The detail structure parameters of $MX_3$ \textit{i.e.}, the lattice constant $a$, the length of $M$-$X$ bond $d(M$-$X)$, the angle between the $M$-$X$ bond and the direction normal to the plane $\theta$, and the experimental lattice constant for the 3D structure  are tabulated in Table~\ref{Tab-Str}. The lattice constants of $MX_3$ monolayers are fully optimized.

\begin{figure}[h]
	\includegraphics[width=12cm]{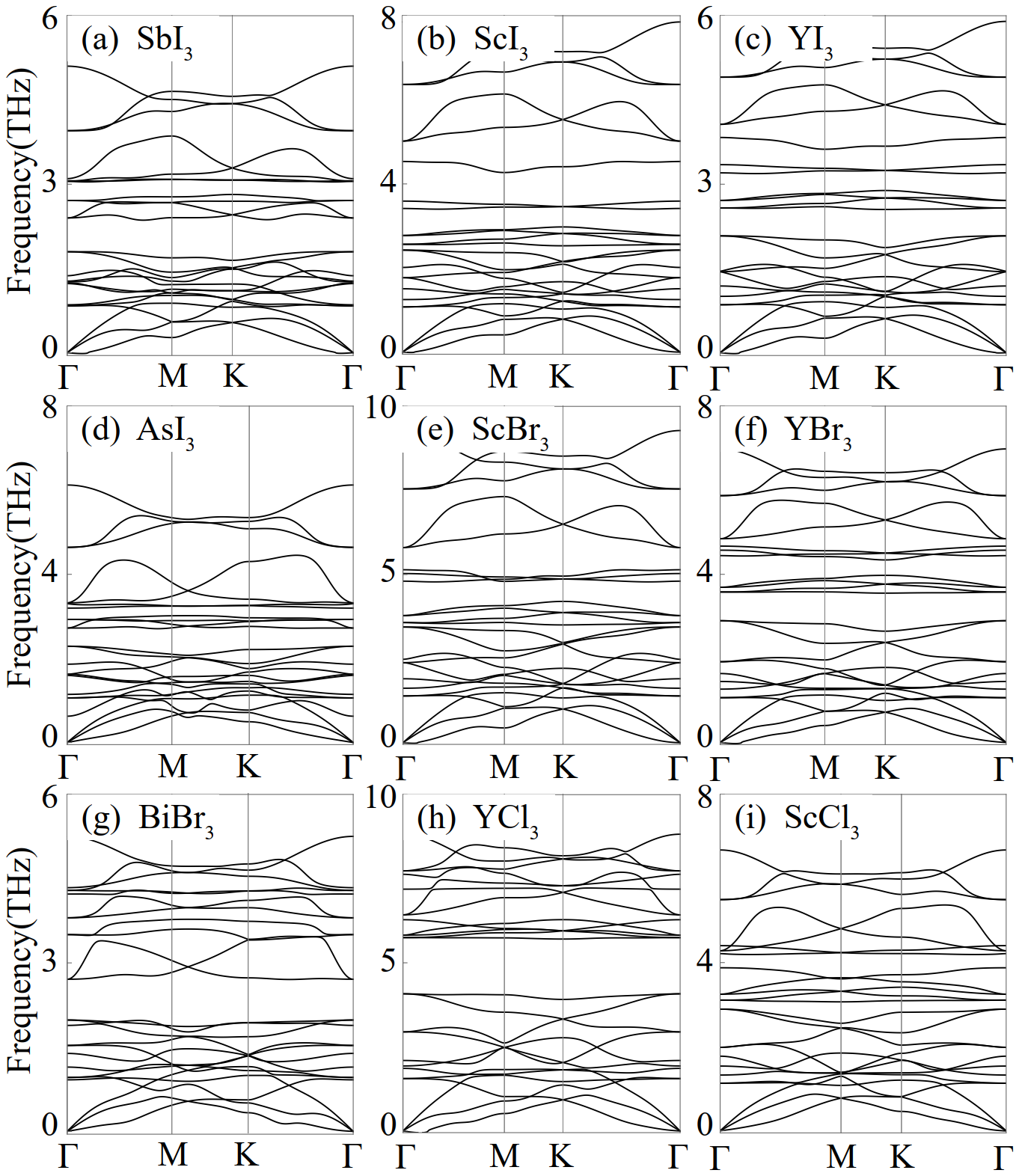}
	\caption{(a)-(i) The calculated phonon spectra of $MX_3$ monolayers. The corresponding computation details are given in Sec.~II. }
\label{figS1}
\end{figure}

Further, with systematically study on the phonon spectra of $MX_3$  ($M$=Bi,Sb,As,Sc,Y; $X$=I,Br,Cl) monolayers, we find there are many gaps that possesses OAI phase in the phonon spectra. 
Here, we show the phonon spectra of  $MX_3$ monolayers in Fig.~\ref{figS1}. The vanish of the imaginary frequency indicate the dynamical stability of $MX_3$ monolayers.
The gaps that are in OAI phase are given in the Table I of the main text.

\begin{table}[h]
\caption{The irreducible representation (irrp) of the phonon branches of the little group at different high-symmetry points in the first BZ of BiI$_3$ monolayer. 
The Irrps are written under the BCS convention.
The number in brackets indicate the dimension of the irrp and the degeneracy fold of the phonon branches at corresponding high-symmetry point. The superscript $+/-$ means the parity. }
\vspace{0.2cm}
\renewcommand\arraystretch{1.6}
\setlength{\tabcolsep}{8mm}{
\begin{tabular}{cccc}
\hline 
\hline 
Level & $\Gamma$ & M & K\tabularnewline
\hline 
1 & $\Gamma_{2}^{-}\left(1\right)\oplus \Gamma_{3}^{-}\left(2\right)$ & M$_{2}^{+}\left(1\right)$ & K$_{3}\left(2\right)$\tabularnewline
2 & $\Gamma_{3}^{+}\left(2\right)$ & M$_{2}^{+}\left(1\right)$ & K$_{1}\left(1\right)$\tabularnewline
3 & $\Gamma_{2}^{-}\left(1\right)$ & M$_{2}^{-}\left(1\right)$ & K$_{2}\left(1\right)$\tabularnewline
4 & $\Gamma_{2}^{+}\left(1\right)$ & M$_{1}^{+}\left(1\right)$ & K$_{3}\left(2\right)$\tabularnewline
5 & $\Gamma_{1}^{+}\left(1\right)$ & M$_{2}^{-}\left(1\right)$ & K$_{3}\left(2\right)$\tabularnewline 
6 & $\Gamma_{3}^{-}\left(2\right)$ & M$_{1}^{-}\left(1\right)$ & K$_{1}\left(1\right)$\tabularnewline
7 & $\Gamma_{3}^{+}\left(2\right)$ & M$_{1}^{+}\left(1\right)$ & K$_{3}\left(2\right)$\tabularnewline
8 & $\Gamma_{2}^{+}\left(1\right)$ & M$_{2}^{-}\left(1\right)$ & K$_{1}\left(1\right)$\tabularnewline
9 & $\Gamma_{3}^{-}\left(2\right)$ & M$_{2}^{+}\left(1\right)$ & K$_{2}\left(1\right)$\tabularnewline
10 & $\Gamma_{3}^{+}\left(2\right)$ & M$_{1}^{-}\left(1\right)$ & K$_{3}\left(2\right)$\tabularnewline
11 & $\Gamma_{3}^{-}\left(2\right)$ & M$_{1}^{+}\left(1\right)$ & K$_{2}\left(1\right)$\tabularnewline
12 & $\Gamma_{1}^{+}\left(1\right)\oplus\Gamma_{1}^{-}\left(1\right)$ & M$_{2}^{-}\left(1\right)$ & K$_{1}\left(1\right)$\tabularnewline
13 & $\Gamma_{3}^{+}\left(2\right)$ & M$_{2}^{+}\left(1\right)$ & K$_{3}\left(2\right)$\tabularnewline
14 & $\Gamma_{2}^{-}\left(1\right)$ & M$_{2}^{-}\left(1\right)$ & K$_{3}\left(2\right)$\tabularnewline
15 &  & M$_{1}^{-}\left(1\right)$ & K$_{3}\left(2\right)$\tabularnewline
16 &  & M$_{2}^{+}\left(1\right)$ & K$_{1}\left(1\right)$\tabularnewline
17 &  & M$_{1}^{+}\left(1\right)$ & \tabularnewline
18 &  & M$_{2}^{-}\left(1\right)$ & \tabularnewline
19 &  & M$_{1}^{-}\left(1\right)$ & \tabularnewline
20 &  & M$_{1}^{+}\left(1\right)$ & \tabularnewline
21 &  & M$_{1}^{-}\left(1\right)$ & \tabularnewline
22 &  & M$_{2}^{+}\left(1\right)$ & \tabularnewline
23 &  & M$_{2}^{-}\left(1\right)$ & \tabularnewline
24 &  & M$_{1}^{+}\left(1\right)$ & \tabularnewline
\hline 
\hline
\end{tabular}}
\label{Tab-BR}
\end{table}

\section{III. The band representation of BiI$_3$}

In this section, we show the irreducible representation of the phonon branches of the little group at different high-symmetry points. The irreducible representations are obtained by the PhononIrep package and written under the BCS convention. The results are  tabulated in Table~\ref{Tab-BR}.

With these irreducible representations, we can get the RSI at different WPs. Based on the TQC theory, with $N_{Occ}=13$, the $\mathbb{Z}$ type RSI for the WP $1a$ is calculated by  
\begin{eqnarray}
\delta_{2}\left(a\right) & = & -m\left(E_{g}\right)+m\left(E_{u}\right)\nonumber \\
 & = & m\left(\Gamma_{1}^{+}\right)-m\left(\Gamma_{1}^{-}\right)+m\left(\Gamma_{2}^{+}\right)+m\left(\Gamma_{3}^{-}\right) -m\left(\mathrm{K}_{1}\right)-m\left(\mathrm{K}_{3}\right)+m\left(\mathrm{M}_{1}^{+}\right)\label{supRSI-1}\\
 & = & 1, \nonumber 
\end{eqnarray}
and the $\mathbb{Z}_2$ type RSI for the WP $6i$ is calculated by  
\begin{eqnarray}
\eta_{4}\left(i\right) & = & \begin{array}{ccc}
\left[-m\left(A\right)+m\left(B\right)\right] & \mathrm{mod} & 2\end{array}\nonumber \\
 & = & \begin{array}{ccc}
\left[m\left(\Gamma_{1}^{+}\right)+m\left(\Gamma_{3}^{+}\right)+m\left(\mathrm{K}_{1}\right)+m\left(\mathrm{M}_{1}^{+}\right)\right] & \mathrm{mod} & 2\end{array}\label{supRSI-2}\\
 & = & 1.\nonumber 
\end{eqnarray}
Here, the function $m$ indicates the number of little group Irrps in the occupied bands. Moreover, in Eqs.~(\ref{supRSI-1}) and~(\ref{supRSI-2}), the fact that the system is in 2D is adopted as a necessary condition. for $N_{Occ}=17$ or $21$ we also get nonzero $\delta_{2}\left(a\right)$ indicating the OAI phase.

\begin{figure}[h]
	\includegraphics[width=5.5cm]{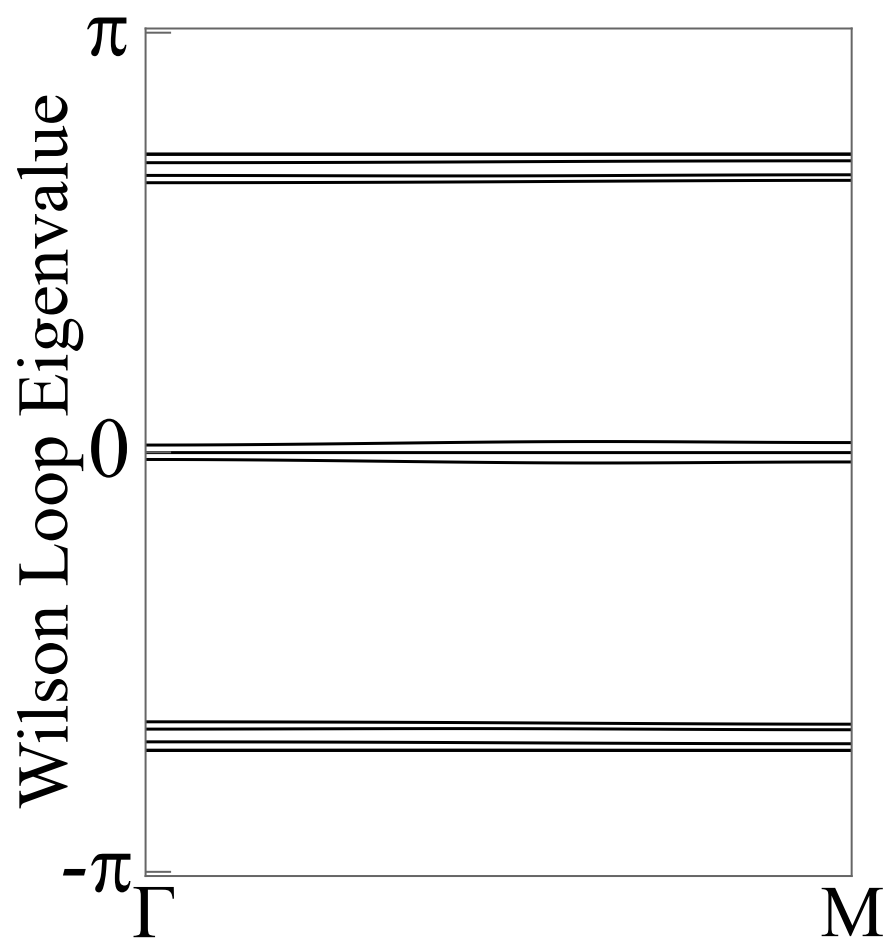}
	\caption{
The wilson loop of of BiI$_3$ monolayer with $N_{Occ}=13$.
    }
\label{figS3}
\end{figure}

Moreover, based on Table~\ref{Tab-BR}, with the number of occupied band is $N_{Occ}=13$, we calculated the eigenvalues of the inversion symmetry. 
We find there is no band inversion, which indicates the nontrivial band topology is beyond the framework of real Chern insulator~\cite{zhao2020equivariant} where floating SS is also reported. 
The further confirmation of this point is provided by the gaped of the wilson loop eigenvalue at $\pi$, as shown in  Fig.~\ref{figS3}.

\section{IV. Corner states with different atom terminals at the edge}

\begin{figure}[h]
	\includegraphics[width=14cm]{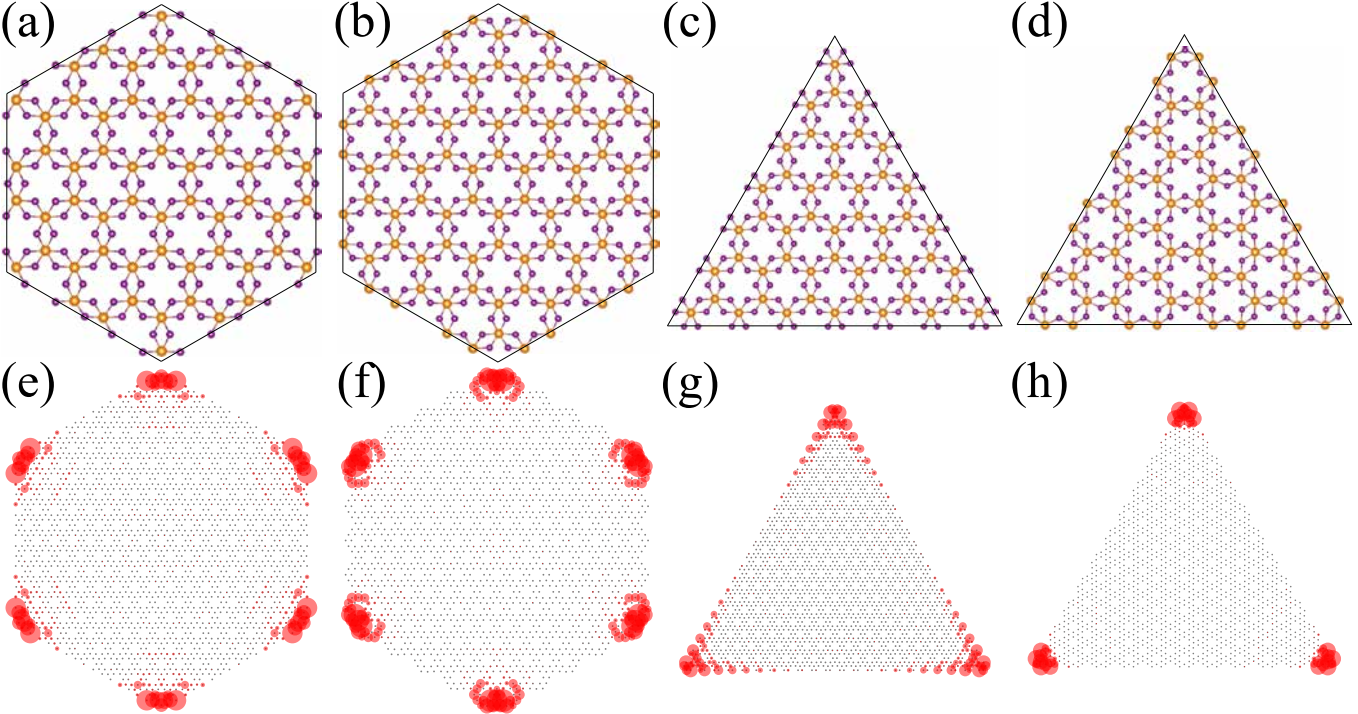}
	\caption{
(a) A  hexagonal-shaped nano-disk cut from Bi$_3$ monolayer with I atoms are exposed at edge.
(b) A  hexagonal-shaped nano-disk of Bi$_3$ monolayer with Bi atoms are exposed at edge. 
(c) A  triangular-shaped nano-disk of Bi$_3$ monolayer with I atoms are exposed at edge.
(d) A  triangular-shaped nano-disk of Bi$_3$ monolayer with Bi atoms are exposed at edge.
(e)-(h) The compute spatial distributions of emerging corner states in the energy spectra for nano-disk shown in (a)-(d).
    }
\label{figS2}
\end{figure}

In this section, we shown the emerging of corner states with different atom terminals at the edge. 
As shown in Fig.~\ref{figS2}, we consider a hexagonal- and triangular-shaped nano-disk. 
In Figs.~\ref{figS2} (a) and (c), the I atoms are at the edges of the nano-disk, while in Figs.~\ref{figS2} (b) and (d) Bi atoms at the edges of the nano-disk. 
One obverses there are many states emerge within the gap formed by $PB_{13}$ and $PB_{14}$. 
We computed the spatial distributions of the states within the bulk gap, the results are shown in Figs.~\ref{figS2} (e)-(h). 
As expected, six (three) fold-degenerated  phonon models that are located at the corner of the nano-disk can be picked out from the phonon spectra of the hexagonal-shaped (triangular-shaped) nano-disk, as shown in Figs.~\ref{figS2} (e)-(h).

\end{document}